\documentclass[aps,rmp,showpacs,superscriptaddress]{revtex4}

\usepackage[latin1]{inputenc}
\usepackage{bm,graphicx,graphics,amsmath,amssymb,epsfig,color}
\usepackage{hyperref}
\setcitestyle{numbers,square,citesep={,}}

\newcommand{\be}{\begin{equation}}
\newcommand{\ee}{\end{equation}}
\newcommand{\bea}{\begin{eqnarray}}
\newcommand{\eea}{\end{eqnarray}}
\newcommand{\bse}{\begin{subequations}}
\newcommand{\ese}{\end{subequations}}
\newcommand{\comment}[1]{}

\begin{document}

\title{Anomalous heat transport in classical many-body systems:\\
overview and perspectives}
\author{G. Benenti}
\email{giuliano.benenti@uninsubria.it}
\affiliation{Center for Nonlinear and Complex Systems, Dipartimento di
Scienza e Alta Tecnologia, Universit\`a degli Studi dell'Insubria,
via Valleggio 11, 22100 Como, Italy}\affiliation{Istituto Nazionale di Fisica Nucleare, Sezione di Milano, via Celoria 16, 20133 Milano, Italy}
\affiliation{NEST, Istituto Nanoscienze-CNR, I-56126 Pisa, Italy}
\author{S. Lepri}
\email{stefano.lepri@isc.cnr.it}
\affiliation{Consiglio Nazionale delle Ricerche, Istituto dei Sistemi Complessi, Via Madonna del Piano 10 I-50019 Sesto Fiorentino, Italy} 
\affiliation{Istituto Nazionale di Fisica Nucleare, Sezione di Firenze, via G. Sansone 1 I-50019, Sesto Fiorentino, Italy}
\author{R. Livi}
\email{roberto.livi@unifi.it}
\affiliation{ Dipartimento di Fisica e Astronomia and CSDC, Universit\`a di Firenze, via G. Sansone 1 I-50019, Sesto Fiorentino, Italy}
\affiliation{Istituto Nazionale di Fisica Nucleare, Sezione di Firenze, via G. Sansone 1 I-50019, Sesto Fiorentino, Italy}
\affiliation{Consiglio Nazionale delle Ricerche, Istituto dei Sistemi Complessi, Via Madonna del Piano 10 I-50019 Sesto Fiorentino, Italy} 

\begin{abstract}
In this review paper we aim at illustrating recent achievements in  anomalous heat diffusion, while highlighting  open
problems and  research perspectives. We briefly recall the main features of the 
phenomenon for low-dimensional classical anharmonic chains 
and outline some recent developments on perturbed integrable systems,
and on the effect of long-range forces and magnetic fields. 
Some selected applications to heat transfer in material 
science at the nanoscale
are described. In the second part, we discuss of the role of 
anomalous conduction on coupled transport and describe how systems 
with anomalous (thermal) diffusion allow a much better power-efficiency
trade-off for the conversion of thermal to particle current. 
%A brief overview of research perspective in quantum systems is given.
\end{abstract}

\pacs{63.10.+a  05.60.-k   44.10.+i}

\maketitle

\section{Introduction}

Anomalous diffusion is a well-established concept in statistical physics and 
has been invoked to describe many diverse kinetic phenomena. A very detailed insight
has been achieved by generalizations of the motion of Brownian particles 
as done for the continuous-time random walk, L\`evy flights and walks.
A formidable body of literature on the topic exists, and we refer for example 
to \cite{klages2008anomalous} as well as the present issue for an overview.

The above particle models are based on a single-particle description, whereby
the single walker performs a non-standard diffusive motion. 
How do these features emerge when dealing with a many-body problem?
What are the conditions for a statistical system composed of many 
interacting particles to yield \textit{effectively} anomalous 
diffusion of particles or quasi-particles? 
A further question regards how such anomalies in diffusion can be 
related to transport and whether they can be somehow exploited to 
achieve some design principle, like efficiency of energy conversion.
In fact, although thermoelectric phenomena are known since centuries,
it is only recently that a novel point of view on the problem
has been undertaken \cite{Benenti2017}. Generally speaking, this renewed research activity
is motivated also by the possibility of  applications of the thermodynamics and statistical mechanics 
to nano and micro-sized systems with applications to molecular biology, micro-mechanics, 
nano-phononics etc. This requires treating systems far from thermodynamic limit,
where fluctuations and interaction with the environment are essentially
relevant and require to be treated in detail.

In the present contribution, we first review how anomalous energy diffusion
arises in lattices of classical oscillator as a joint effect of nonlinear 
forces and reduced dimensionality (and in this respect we will 
mostly discuss one-dimensional chains). In words, this amounts to
say that the anomalous dynamics of energy carriers is an emergent
feature stemming from correlations of the full many-body dynamics.
As a consequence, Fourier's law \textit{breaks down}: motion of energy carriers is so
correlated that they are able to propagate \textit{faster} than diffusively. 
In the second part of the paper, we discuss how this feature influences
coupled transport and how it can be used to enhance the efficiency
of thermodiffusive processes.

We conclude this review about the multifaceted problem of heat transport in classical systems with a short 
overview summarizing possible extensions to the quantum domain, with reference to related open problems,
which certainly will deserve attention in the near future.

\section{Anomalous heat transport in classical anharmonic lattices} 
\label{sec:anomal}

The presence of a diverging heat conductivity with the systems size in a chain of
coupled nonlinear oscillators was first pointed out in \cite{LLP97,Lepri98a}. This was the
beginning of a research field that, over more than two decades,  has been devoted 
to the understanding of the mechanisms yielding anomalous transport in low-dimensional systems.
Far from being a purely academic problem, it has unveiled the possibility of observing
such peculiar effects in nanomaterials, like nanotubes, nanowires or 
graphene \citep{Chang2016,Balandin2011}.
Extended review articles on this problem exist since many years 
\cite{LLP03,DHARREV}, while a collection of contributions about more recent achievements is
contained in \cite{Lepri2016} and in  
a review article  of the present issue \cite{Dhar2019}.  
Here it is useful to provide first a short summary
about the state of the art in this field, while the main effort will be devoted to 
some recent achievements that point to promising and challenging directions for future investigations.

In any model where anomalous transport has been 
observed it emerges as a hydrodynamic effect due to the conspiracy of reduced 
space-dimensionality and conservation laws, yielding non standard relaxation properties even in
a linear response regime. As a reference we may consider the basic class of models,
represented by a Hamiltonian of the following form:
\begin{equation}
H = \sum_{n=1}^L \left[\frac{p_n^2}{2m} + V(q_{n+1} - q_n) \right]\,.
\label{Hamil}
\end{equation}
Typical choices for the interaction is the famous Fermi-Pasta-Ulam-Tsingou (FPUT) potential 
where $V(x) = V_{FPUT}(x) \equiv \frac{1}{2} x^2+\frac{\alpha}{3} x^3 + \frac{\beta}{4} x^4$
and rotor (or Hamiltonian XY) model $V(x)= V_{XY}(x) \equiv 1-\cos x$.
For what concerns conservation laws, in $d=1$ anomalous transport has been
generically observed in Hamiltonian models  of type (\ref{Hamil}), where energy and momentum and the "stretch"
variable $\sum_n(q_{n+1}-q_n)$ are conserved.  It is worth recalling that any  approach aiming at
describing out-of-equilibrium conditions, like stationary 
transport processes, has to be based on the hydrodynamic equations associated with such locally conserved quantities.
One-dimensional oscillator models
with only one (e.g. the Frenkel-Kontorova or $\phi^4$ models \cite{Aoki00}) or two 
conserved quantities like the rotor model \cite{Giardina99,Gendelman2000} or the 
discrete nonlinear Schr\"odinger lattice \cite{Iubini2012,Mendl2015} show instead
standard diffusive transport. 
 Intuitively, this is due to the presence of scattering sources for acoustic waves
propagating through the lattice imposed by the very presence of a local nonlinear
potential, which breaks translation invariance (i.e. momentum conservation). This argument does not
apply to the rotor model, where stretch only is not conserved, due to the  angular nature of the 
$q_n$ variables: anyway standard diffusion is allowed because of the boundedness of the cosine
potential.
For what concerns dimensionality, in $d=3$ 
normal diffusion regimes are expected to characterize heat transport
 in nonlinear lattices.
%, with the exception of integrable models like a purely harmonic solid. 
Only in $d=2$ one can find evidence of a diverging heat conductivity according to a 
logarithmic dependence with the system size $L$ \cite{Lippi00,Wang2012,DiCintio2017}.

\begin{figure}
\includegraphics[width=0.8\textwidth]{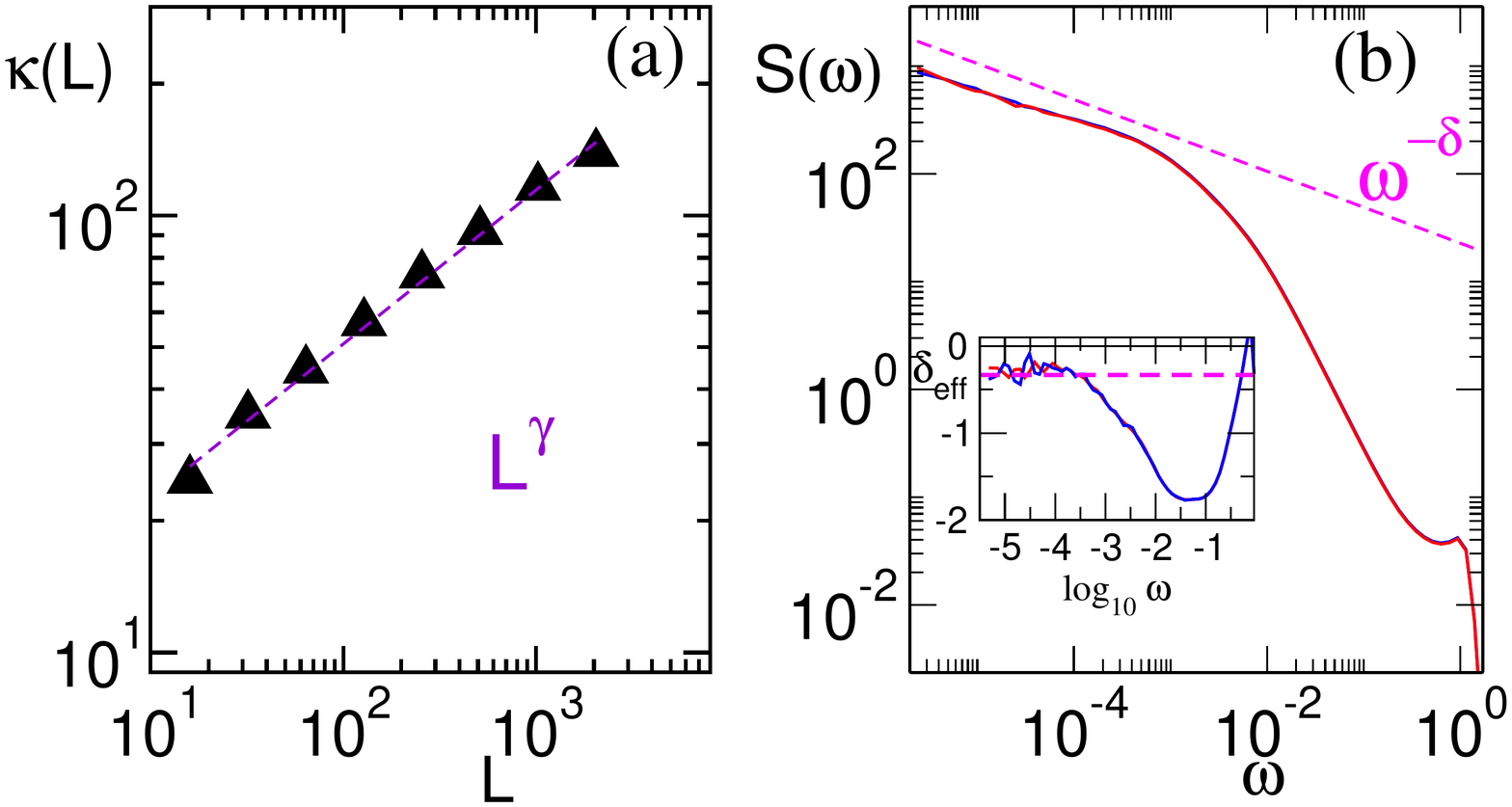}
\caption{Anomalous thermal conductivity for the FPUT model 
with cubic and quartic potential term ($\alpha=0.25$, $\beta=1$); (a) finite-size conductivity measured in the nonequilibrium steady state and  (b) power spectrum of heat current
fluctuations $S(\omega)$ (i.e. the Fourier transform of $\langle J(t)J(0)\rangle$); 
the long-time tail (\ref{longtail}) in (\ref{longtail}) corresponds to a divergence 
$\omega^{-\delta}$ at small frequencies. The inset reports the logarithmic derivative
$\delta_{eff}=d\log_{10} S/d\log_{10} \omega$. Data are  compatible with $\delta=\gamma=1/3$
(dashed lines). Microcanonical simulations at energy density $0.5$.
% energy 0.5 alpha=0.25
}
\label{fig:anomal}
\end{figure}

The main distinctive feature of anomalous heat transport in one-dimensional Hamiltonian
models of anharmonic lattices is that
the finite-size heat conductivity $\kappa(L)$  diverges  in
the limit of a large system size $L\to \infty$ \cite{LLP97} as
\[
\kappa(L) \;\propto\; L^\gamma \,,
\]
with $0 < \gamma \le1$,  (the case $\gamma = 1$ corresponding to integrable 
models, like the Toda lattice discussed in Sec. \ref{sec:chim}).
This implies that
this transport coefficient is,in the thermodynamic limit,  not well-defined. 
In the linear response
regime, this is equivalent to find that 
the equilibrium correlator of the
energy current $J(t)$ displays for long times $t$ a nonintegrable power-law decay, 
\begin{equation}
\langle J(t)J(0)\rangle \propto t^{-(1-\delta)} \,,
\label{longtail}
\end{equation}
with $0\le\delta < 1$. Accordingly, 
the Green-Kubo formula yields an infinite value of the heat conductivity
and allows to establish the equivalence of the exponents, i.e.  $\gamma = \delta$, 
 provided sound velocity is finite  \cite{Lepri98a}.
In Fig.\ref{fig:anomal} we illustrate two typical simulations of the 
FPUT model demonstrating the results above.

%%%%%%%%%%%%%%%%%
%%%%%%%%%%%%%%%%%

The most basic issue of anomalous feature is related to anomalous dynamical scaling of equilibrium correlation of the hydrodynamic modes. A simple way to state this is to say that fluctuations of the conserved quantities with small wavenumber $k$ evolve
on time scales of order $\tau(k) \sim |k|^{-z}$. For standard diffusion one has $z=2$.
Within the nonlinear fluctuating hydrodynamics approach it has been 
shown \cite{VanBeijeren2012,Spohn2014} that models like 
(\ref{Hamil}) belong generically to the universality class 
of the famous Kardar-Parisi-Zhang
(KPZ) equation, originally formulated in the context of growing interfaces. 
It is well known that this equation in $d=1$ is characterized by the dynamical exponent $z=3/2$. 
The origin of this nontrivial dynamical exponent can be traced back to the 
nonlinear interaction of long-wavelength modes. This leads to the prediction
$\gamma= (2-z)/z=1/3$ (at least in the linear response regime), a value that should 
be largely universal, as confirmed by many numerical experiments.

The above consideration applies generically for anharmonic chains 
with three conservation laws \cite{Spohn2014}.
There is however the possibility of having a different
universality class depending on the number of conserved quantities \cite{Popkov2015} and/or on the 
nonlinear coupling among the hydrodynamic modes \cite{Spohn2014}. For, instance model 
(\ref{Hamil}) with an even potential $V(x)=V(-x)$ should belong to the 
a different universality class having a different exponent $\gamma$.
Actually, the precise value $\gamma$ is still somehow controversial:
the theoretical prediction from mode-coupling approximation of 
the hydrodynamic theory yields $\gamma= 1/2$ \cite{Lee-Dadswell05,Delfini07b,Spohn2014}
while kinetic theory yields $\gamma= 2/5$ \cite{Lukkarinen2008},
a value closer to the one measured numerically \cite{Lepri03,Wang2011}
(see also \cite{BBO06,Lepri2009} for related results on exactly solvable
models). The existence of the two classes can be demonstrated
either by direct measurements of the exponents \cite{Lepri03}, but also by suitable changes of the 
thermodynamic parameters. For instance, a nonlinear chain with a symmetric 
potential subject to a suitable pressure acting at its
boundaries may exhibit a change from exponent $\gamma = {1}/{2}$ to 
$\gamma = 1/3$ \cite{Lee-Dadswell2015}. 
%Citerei i lavori di Lee-Dadswell
This is a relevant observation for possible experimental verification of anomalous
heat transport, because the pressure or torque applied to 
any one-dimensional material should be taken into account for a correct comparison
with theoretical predictions.

A physically intuitive way to describe anomalous heat transport is to think 
in terms of a L\'evy walk, namely an ensemble of random walkers performing 
free ballistic steps with finite velocity for times that are power-law distributed \cite{Zaburdaev2015}.
This simple description accounts very well for many features of anharmonic lattices
and fluids in various non-equilibrium setups \cite{Cipriani05,Lepri2011,Dhar2013}.    
For instance  
energy perturbations propagate superdiffusively \cite{Denisov03,Cipriani05,Liu2014}: 
an initially localized perturbation of the energy broadens and its variance grows
in time as $\sigma^2 \propto t^{\eta}$ with $\eta > 1$. 
These empirical observation have a theoretical justification 
within the nonlinear fluctuating hydrodynamics. Indeed the theory
predicts a hydrodynamic "heat mode" that has the characteristic 
shape given by a L\'evy-stable distribution, see \cite{Spohn2014,Mendl2014} for details.
Further support comes from mathematical results: superdiffusive 
behavior has been proven for one dimensional infinite chain of harmonic oscillators undergoing stochastic collisions conserving energy and 
momentum \cite{Jara2015,Bernardin2016}. 
 In the same spirit, the more difficult case 
of nonlinear oscillators with conservative noise has been 
discussed \cite{Bernardin2012}. For exponential interactions
(the Kac-van Moerbecke model), 
superdiffusion of energy is again
demonstrated, and a lower bound on the decay of current correlation 
function is given \cite{Bernardin2014}. In Ref. \citep{Spohn2014}
it is argued also such models should belong to the KPZ class.

One related distinctive feature of anomalous transport is that the temperature profiles in nonequilibrium steady states  are nonlinear, even for vanishing applied temperature gradients \cite{Delfini10,Lepri2011}. There is indeed a close connection with fractional heat equation, that has been demonstrated and discussed in the recent literature  \cite{Kundu2019,Dhar2019}.

%\subsection{Perspectives}
%\begin{itemize}

%\item Molecular dynamics polymers and nanowires \cite{Colombo2018,Upadhyaya2016} 

%\item To be understood \cite{Xiong2018} \cite{Hurtado2016}

%\item Transport in Long-range models \cite{Bagchi} \cite{Detal}

%\item Transport in magnetic field \cite{Saitomag}

%\item Local Equilibrium ? (Rondoni)
%\end{itemize}
\subsection{The importance of being small}
\label{sec:small}
As mentioned in the previous section, theoretical inferences on the problem of heat transport in
anharmonic chains stems from the basic assumption that one should compute
any relevant quantity in the limits  $L\to \infty$ and $t\to \infty$,
performed just in this order. On the other hand, in any numerical simulation 
or in real low-dimensional heat conductors, such as nanowires, carbon nanotubes,
polymers or even thin fibers, we have to deal with finite size and finite time corrections.
These can be taken under control in a linear response regime if 
the mean-free path of propagating excitations, $\lambda$, and their mean interaction time, $\tau$,
are such that $\lambda \ll L$ and $\tau \ll t$. It is a matter of fact that, when dealing with 
models of anharmonic chains, such a control is often not granted, mainly because of nonlinear
effects.  This is may be a very relevant problem also for interpreting 
possible experimental verifications of anomalous transport in real systems and also for designing 
nanomaterials, exhibiting deviations from the standard diffusive conductivity. 

 As a matter of fact, very severe finite-size effects 
invariably arise when trying to check predictions numerically. Very often, the 
estimate of the relevant exponents $\gamma$ or $\delta$ systematically
deviates from the expected values, and sometimes are even claimed 
to depend on parameters \cite{Iacobucci2010,Hurtado2016,Xiong2016,Barik2019}. 
If universality is (as we do believe)  to hold, this effects should 
be due to subleading corrective terms to the asymptotics 
that are still relevant
on the scales accessible to simulations. 
Besides those issues, other unexpected effects 
arise. For instance, for 
the FPUT \cite{Basile08}, Toda \cite{Iacobucci2010} and the
Kac-van Moerbecke \cite{Bernardin2016} chains perturbed by conservative noise,  
the exponent $\gamma$ increases with the noise strength. 
Besides the problem of evaluating the precise 
exponents, this is pretty surprising since it suggest that 
a larger stochasticity in the model makes the system 
more diffusive, at least for finite systems.

Another instance where finite size corrections are ``amplified'' by nonlinear effects is the case of anharmonic chains with asymmetric potential , i.e. with $V(x) \not= V(-x)$ as the  FPUT model with $\alpha\neq0$.
As already shown in Fig.\ref{fig:anomal}, 
both equilibrium and out of equilibrium measurements of the heat conductivity in the presence of an 
applied thermal gradient are usually consistent with KPZ scaling.
However, in other temperature regimes, 
Fourier's law appears to hold, i.e. thermal conductivity is 
constant over a large range of sizes \cite{Zhong2012}.
This was traced back  to  the relatively long relaxation time of mass inhomogeneities induced by the asymmetry of the
interaction potential and acting as scatterers of
phonons \cite{Zhong2012}. Actually, it was 
later shown in \cite{Wang2013, Das2014, Chen2014}
that this is a strong finite-size effect, since it persists for relatively large values of $L$ and $t$. Yet,
the expected theoretical prediction of a diverging heat conductivity can be recovered in
simulations performed for sufficiently large values of $L$ and $t$.  
It should be pointed out that all of these speculations rely on numerical results, while a theoretical approach capable to provide estimates of the combination of nonlinear with finite-size
corrections to the hydrodynamics would be useful. 
Indeed, fluctuating hydrodynamics provide some 
prediction: subleading corrections to the leading asymptotic decay 
Eq. (\ref{longtail}) can be very large and decay very slowly \cite{Spohn2014}.

 Finite size-effects can have other facets. A remarkable 
example is the Discrete Nonlinear Schroedinger equation, a well-known 
model for atomic condensates in periodic optical lattices. The 
model has two conserved quantities (energy and number density) 
and display normal diffusive transport \citep{Iubini2012}.
However, at very low temperatures there appears a further almost
conserved quantity (the phase difference among oscillators) 
and, for a finite chain and long times the dynamics is the the same of 
a generic anharmonic model, leading to KPZ scaling of correlations
and anomalous transport \cite{spohn2014fluctuating,Kulkarni2015}.
Further unexpected features have been reported also
in \cite{Xiong2018}. In this paper the authors study this problem for a the FPUT-$\beta$ 
model (i.e. Eq. (\ref{Hamil}) with $V=V_{FPUT}$ and $\alpha=0$)
with an additional local, also called "pinning", potential of the form
\begin{equation}
U({q}) = \frac{1}{2} \sum_{n=1}^L q_n^2 \,.
\label{Pinn}
\end{equation}
This term breaks translational invariance making energy the sole conserved quantity.
By varying the nonlinear coupling $\beta$ one observes a crossover from a ballistic
transport, typical of an integrable model,  to an anomalous diffusive regime ruled by
and exponent of the time correlation function, which
corresponds to a value of $\gamma \sim 0.2$. The crossover occurs in the parameter 
region $0.1 < \beta <1$. Numerical simulations performed for
a chain of a few thousands of oscillators show that further increasing $\beta$ seems to
yield an increasing $\gamma$. The overall outcome challenges the basic theoretical
argument, which predicts that an anharmonic chain equipped by a local potential
should exhibit normal diffusion. For the sake of completeness it is worth mentioning that
in this paper the model under scrutiny is compared with the so-called $\phi^4$-model,
where the nonlinear term in Hamiltonian (\ref{Hamil}),  i.e. $\beta  (q_{n+1} - q_n)^4$, is substituted with $\beta  q_n^4$. Also for this model one observes,  in the same parameter region, a crossover
from a ballistic regime to an anomalous diffusive regime, but for $\beta > 1$   
one eventually obtains numerical estimates yielding $\gamma \sim 0$,
i.e. the expected diffusive behavior is recovered.

 All of these results have a logical interpretation only if we
 invoke, once again, the role of finite size corrections combined with nonlinearity.
 Actually, in the $\phi^4$-model there is no way to argue that a ballistic regime
 should be observed for any finite, even if small, value of $\beta$.  The ballistic
 behavior observed in both models for $\beta < 0.1$ seems to suggest that for small
 nonlinearities one needs to explore definitely much larger
 chains and integrate the dynamics over much longer times, than those employed in \cite{Xiong2018},
 before in both chains phonon-like waves may experience the
 scattering effects due to the local potential. 
 Moreover, the weaker quadratic pinning potential
 of the original model seems to be still affected by finite size corrections, even  in the
 region $\beta >1$.  A problem that one should investigate systematically is the dependence
 on $\beta$ of the chain length and of the integration time necessary to recover standard
 diffusive transport, at least in the crossover region $0.1 < \beta <1$, where one can
 expect to perform proper numerical analysis in an accessible computational time.

\subsection{Chimeras of ballistic regimes}
\label{sec:chim}
In the light of what discussed in the previous section one should not be surprised to 
encounter further  nonlinear chain models, equipped with a pinning potential, which 
exhibit a regime of ballisitic transport of energy, compatible with a linearly divergent 
heat conductivity, $\kappa(L) \sim L$.
Again, one could conjecture that this is due to the  puzzling 
combination of nonlinearity and finite size effects, although, as we are going to
discuss, the emerging scenario is more intricate and interesting than the former
statement could foretell.

As a preliminary remark we want to recall that ballistic transport is the typical situation of an integrable Hamiltonian chain, whose prototypical example is the harmonic lattice, with
$V(x) = \frac{1}{2} x^2$ in Hamiltonian (\ref{Hamil}). 
It is worth pointing out that the addition of the harmonic pinning term
(\ref{Pinn}) again keeps the harmonic chain integrable.
%Qui ci vuole una referenza
 In this perspective, it seems
reasonable to find that for sufficiently small nonlinearities both the FPUT-$\beta$ 
and the $\phi^4$ chains, as discussed in the Section \ref{sec:chim}, may exhibit a seemingly ballistic regime for $\beta < 0.1$
also in the presence of the pinning potential. Recovering the expected diffusive transport regime
is a matter of simulating exceedingly large chains over exceedingly long times.  

Anyway, the peculiar role played by the quadratic pinning potential  (\ref{Pinn}) has emerged also
in a recent study of heat transport in the Toda chain \cite{DiCintio2018}. It is worth recalling that the unpinned
Toda chain  is an integrable Hamiltonian model of the form (\ref{Hamil}) with $
V(x)= e^{-x} + x -1 $: in this model heat transport is ballistic
due to the finite--speed propagation of solitons (rather than phonons, as in the  
harmonic chain). Toda solitons are localized nonlinear excitations
which are known to interact with each other by a non--dissipative diffusion
mechanism: the soliton experiences a random sequence of spatial shifts 
while it moves through the lattice 
and interacts with other excitations without exchanging momentum  \cite{Theod1999}.
%Theodorakopoulos N, Peyrard M. Solitons and nondissipative diffusion. Phys Rev Lett 1999;83(12):2293.
In fact, the calculation of the transport coefficients by the Green-Kubo formula indicates the
presence of a finite Onsager coefficient, which corresponds to a diffusive process on top of
the dominant ballistic one \cite{Spohn2018, Kundu2016}.
% \cite{Spohn2018},\cite{Kundu2016}).
%Spohn H. Interacting and noninteracting integrable systems. J Math Phys 2018;59(9):091402. 
%Kundu A, Dhar A. Equilibrium dynamical correlations in the Toda chain and other integrable models. Phys Rev E 2016;94:062130.

When the pinning term (\ref{Pinn}) is at work the Toda chain becomes chaotic, as one can easily
conclude by measuring the spectrum of  Lyapunov exponents \cite{DiCintio2018}.
This notwithstanding, not only the energy, but also the "center of mass'' 
$$
h_c= \frac{1}{2} (\sum_{n=1}^L q_n)^2 + \frac{1}{2}(\sum_{n=1}^L p_n)^2
$$
is a conserved quantity. The peculiarity of the quadratic pinning potential (\ref{Pinn}) is testified by the
observation that if one turns it into a quartic one, i.e. $U(q_i) = \frac{1}{4} \sum_{n=1}^L q_n^4$,
the quantity $h_c$ is no more conserved. 
Nonequilibrium simulations of the Toda chain with the addition of (\ref{Pinn}), where heat reservoirs
at different temperature, $T_1 > T_2$ act at its boundaries, yield a "flat" temperature profile at
$T = (T_1 + T_2)/2$ 
in the bulk of the chain. This scenario would be expected for the pure Toda chain,
but it is inconsistent with the basic consideration that the presence of  (\ref{Pinn}) breaks translation
invariance and total momentum is no more conserved. This notwithstanding, in order to observe a 
temperature profile in the form of a linear interpolation between $T_1$ and $T_2$ (i.e., Fourier's law)
one has to simulate the dynamics of very large chains over very long times:
typically $L \sim {\mathcal O}(10^4)$ and $t  \sim {\mathcal O}(10^6)$, when all the parameters of the
model are set to unit.

Equilibrium measurements based on the Green-Kubo relation, i.e. on the 
behavior of the energy current correlator (\ref{longtail}), provide further interesting
facets of this scenario.  By comparing the Toda chain with quadratic and quartic
pinning potentials one observes in the latter case clear indications of a diffusive regime, i.e.  a finite
heat conductivity, and a practically negligible influence
of finite size corrections,
while in the former case the power spectrum (i.e. the Fourier transform of (\ref{longtail})) is found to exhibit 
a peculiar scaling regime (with a power $-5/3$), before eventually reaching a plateau,
that indicates a standard diffusion. In the same region of the spectrum the 
FPUT model (where the parameters $\alpha$ and 
$\beta$ have been chosen in such a way to correspond to a Taylor series expansion
of the Toda chain) with the addition of (\ref{Pinn}) is found to converge to a plateau,
in the absence of any precursor of a power-law scaling.

Further details about the unexpected transport regimes
encountered in the Toda chain equipped with the quadratic pinning can be found in
\cite{Dhar2018}.
%Pierfrancesco Di Cintio et al. "Transport in perturbed classical integrable systems: The pinned Toda chain"
%Chaos, Solitons and Fractals 117 (2018) 249???254 
%and
%Dhar, Kundu, Lebowitz, Scaramazza arXiv:1812.11770
One should point out that many of them are still waiting for a convincing
theoretical interpretation.

\subsection{Anomalous transport in the presence of a magnetic field}
\label{sec:magn}
Recent contributions \cite{Tamaki2017,Saito2018}
%Shuji Tamaki, Makiko Sasada, and Keiji Saito, PRL 119, 110602 (2017)
%Keiji Saito and Makiko Sasada, Commun. Math. Phys. 361, 951???995 (2018)
have tackled the important problem of heat transport in chains of charged oscillators 
in the presence of a magnetic field ${\bf B}$. The model at hand is a one-dimensional polymer, that
allows for transverse motion of the oscillators interacting via a harmonic potential. For ${\bf B} = 0$
the exponent of the energy current correlator is $\delta = \frac{1}{2}$, thus indicating the presence
of anomalous transport and a divergent heat conductivity $\kappa (L) \sim L^{\frac{1}{2}}$.
By switching on ${\bf B}$ the first basic consequence is the breaking of translation invariance, so
that the total momentum is no more conserved. This notwithstanding, the total pseudo-momentum is
conserved, but the hydrodynamics of the model is definitely modified. Actually, numerical and 
analytic estimates indicate that the exponent $\delta$ may turn to a value different from $\frac{1}{2}$.
In particular, in  \cite{Tamaki2017}
%Shuji Tamaki, Makiko Sasada, and Keiji Saito, PRL 119, 110602 (2017)
two different cases were considered: the one where oscillators have the same charge
and the one where oscillators have alternate charges of sign $(-1)^n$, $n$ being  the integer 
index numbering oscillators along the chain. It can be easily shown that in the former case
the sound velocity is null and the energy correlator exhibits a thermal peak centered  at the origin
and spreading in time. Conversely, in the latter case the sound velocity 
has a finite, ${\bf B}$-dependent value
and the thermal peak of the energy correlator is coupled to sound modes
propagating through the chain. 
In the case of finite sound velocity (alternate charges)
the exponent ruling the divergence of the heat conductivity with the system
size is found to remain the same obtained for ${\bf B} = 0$, i.e. $\gamma = \frac{1}{2}$.
This is not surprising, since also for ${\bf B} = 0$ the sound velocity in the model is finite. 
In the case of equally charged bids, on the other hand, it appears a new exponent $\gamma =\frac{3}{8}$, which
corresponds to a universality class, different from all the others encountered  in anomalous transport
in nonlinear chains of oscillators. An important remark on this new exponent is that, in the absence of 
a finite sound velocity, the identification of the exponents $\delta$ and $\gamma$, introduced in Sec.\ref{sec:anomal},
is no more correct. In fact, in this case the value of the exponent $\delta$ is found to be very close to $\frac{3}{4}$.
Rigorous estimates of all of these exponents and also for the $d=2$ and $d=3$ versions of the charged polymer model
have been obtained through the asymptotics of the corresponding Green-Kubo integrals, where the
deterministic dynamics has been  substituted with a stochastic version that conserves the same quantities \cite{Saito2018} . 
%Keiji Saito and Makiko Sasada, Commun. Math. Phys. 361, 951???995 (2018)
For what concerns the different one--dimensional cases, these rigorous estimates  agree with the previous
findings, while in $d=2$  and in $d=3$ the expected logarithmic divergence and a finite heat conductivity have
been singled out, respectively.

\subsection{The case of long-range interactions}
\label{sec:long}

Long-range forces, slowly decaying in the relative distance between particles
are well-studied in statistical mechanics.   They characterize  a wide domain of physical 
situations, e.g.  self-gravitating systems, plasmas, interacting vortices in fluids, capillary effects of colloids at an
interface,  chemo-attractant dynamics, cold atoms in optical lattices, colloidal active particles etc. 
Several unusual features are known: ensemble inequivalence,
long-living metastable states and anomalous energy diffusion \cite{Bouchet2010,RuffoRev}, 
inhomogeneous stationary states \cite{gupta2016surprises},
lack of thermalization 
on interaction with a single external bath \cite{deBuyl2013}, etc. 
Moreover perturbations can spread with infinite velocities, 
leading to qualitative difference with respect to their short-ranged 
counterparts \cite{Torcini1997,Metivier2014}.

The study of heat transport in chains with long-range interactions has been tackled only
recently \cite{avila2015length,Olivares2016,Bagchi2017,tamaki2019energy,wang2019extremely}.
The main question is to what extent are the anomalous properties changed when
the spatial range of interaction among oscillators increases.
In two recent papers \cite{Iubini2018,DiCintio2019} 
this problem has been investigated for Hamiltonian chains with
a long range potential of the form
\be
\label{eq:int_pot}
V = \frac{1}{2N_0(\alpha)}  \sum_{i=1}^{N} \sum_{i\neq j}^N \frac{v(q_i-q_j)}{|i-j|^\alpha}  \, ,
\ee
where the generalized Kac factor
\be
\label{eq:kac}
N_0(\alpha) = \frac{1}{N} \sum_{i=1}^N\sum_{j\neq i}^N \frac{1}{|i-j|^\alpha} \, .
\ee
guarantees the extensivity of the Hamiltonian \cite{Bouchet2010,RuffoRev}. 
In particular, the long-range versions of both
the rotor chain $v(x)=V_{XY}(x)$
and the FPUT-$\beta$ model, $v(x)= \frac{1}{2}x^2+\frac{1}{4}x^4$
have been investigated.
The reason of this twofold choice is that their nearest-neighbor versions (i.e., $\alpha \to +\infty$) of the former model exhibits standard diffusion of energy, while the latter is characterized by anomalous diffusion. 

\begin{figure}
\includegraphics[width=0.55\textwidth]{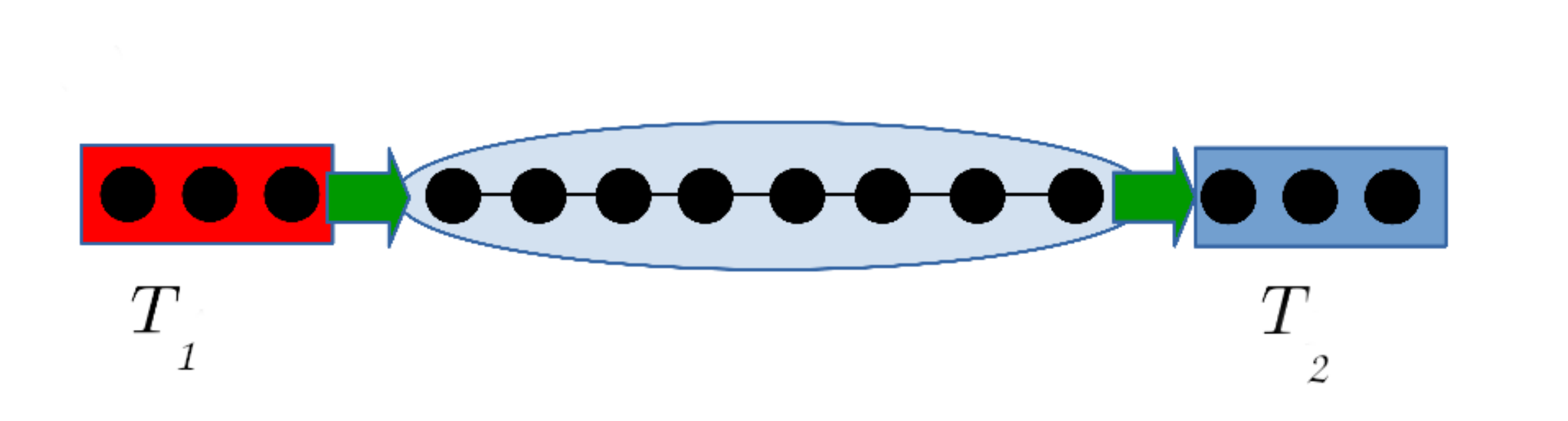}
\includegraphics[width=0.3\textwidth]{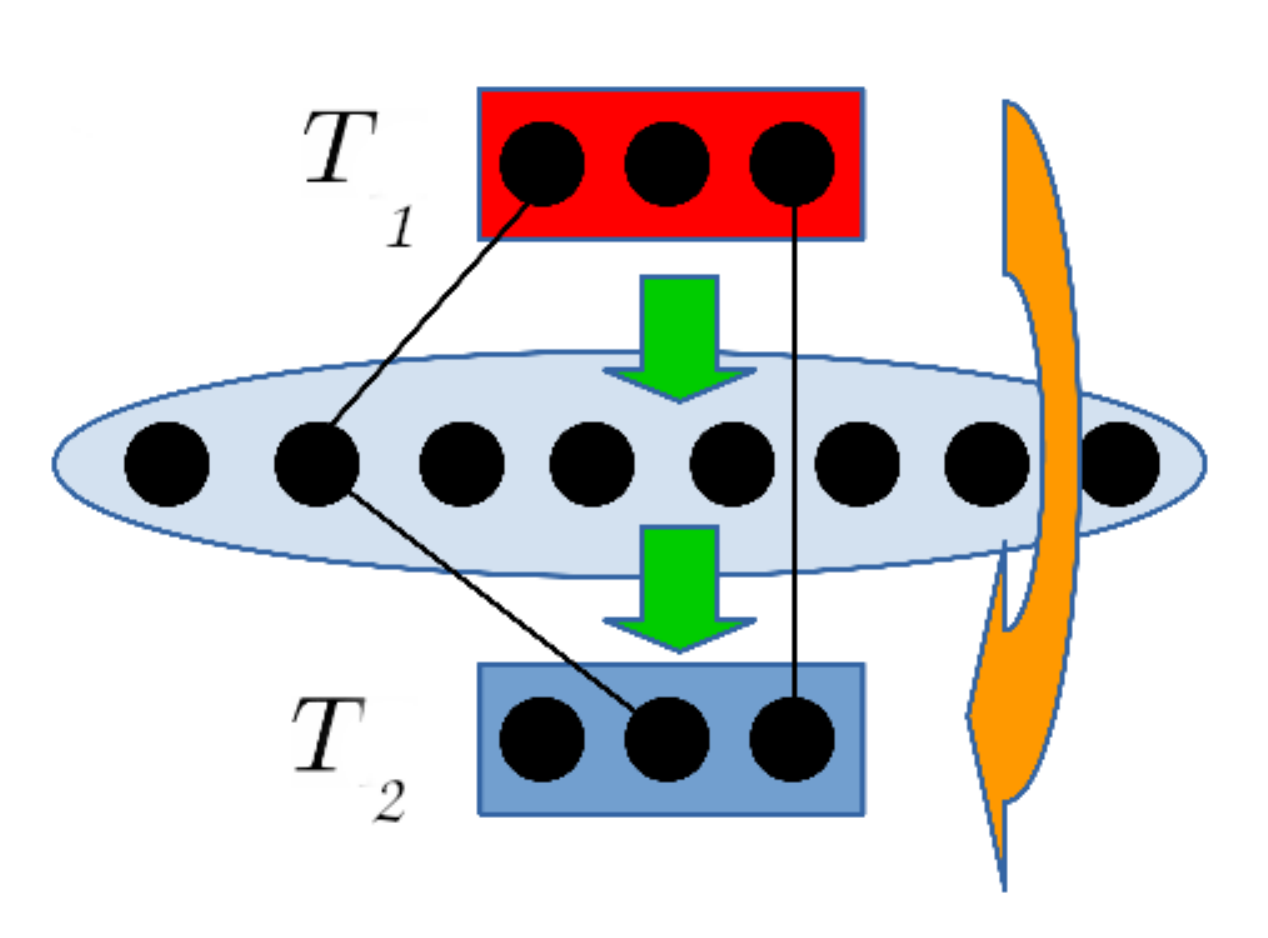}
\caption{A pictorial representations of heat transfer processes for long-range 
interacting chain, in the limit 
case $\alpha =\infty$ (left) and $\alpha=0$ (right).  Oscillators 
in contact with thermal reservoirs are contained in the rectangular boxes while the bulk ones are represented in the ellipse. The relevant transport channels are represented by black lines (adapted with permission from \cite{DiCintio2019}  Copyright (2019) by American Physical Society).}
\label{fig:schema}
\end{figure}

For what concerns the rotor chain, nonequilibrium measurements, with thermal
reservoirs at different temperature, $T_1 > T_2$, acting at the chain ends, show that when $\alpha >1$, the 
resulting temperature profile interpolates linearly between $T_1$ and $T_2$. Despite the long range nature
of the interaction this is a strong indication that a standard diffusive process still governs energy transport
through the rotor chain, as in the limit $\alpha \to +\infty$.  Conversely, when $\alpha < 1$ the temperature profile progressively flattens,
until reaching a constant bulk temperature $T = (T_1 + T_2)/2$ in the "mean-field" limit $\alpha \to 0^+$. It is important to point out that, at variance
with the situation of chimeras of integrable models discussed in Section \ref{sec:chim}, such flat temperature
profiles have nothing to do with integrability, but they are driven by the dominance of a "parallel" energy transport
mechanism, that connects the heat  baths at the chain boundaries with each other directly through each individual
rotors in the bulk of the chain. Energy transport along the chain is practically immaterial and the overall process
is mediated by rotors, that have to compromise between the two different temperatures imposed by the
reservoirs. A sketch of this mechanism is presented in figure \ref{fig:schema}.
For small $\alpha$ each lattice site in the bulk is directly coupled to both thermal baths
and its temperature sets to the average $(T_1+T_2)/2$ independently of the system size.
Moreover, the average heat current exchanged with any other site in negligibly small.
%S. Iubini et al. , PHYSICAL REVIEW E 97, 032102 (2018)
 
At least for $\alpha < 1$, a similar scenario seems to characterize the FPUT model : flat temperature profiles are observed also in
this case and one can verify that the same parallel transport mechanism described  for rotors is at work.
On the other hand, the scenario is definitely more complicated for $\alpha > 1$.
Careful numerical studies, exploring  finite size effects, indicate an overall scenario where an anomalous
diffusion mechanism sets in, characterized by an exponent $\gamma$, that is expected to increase up to the one of the
quartic FPUT model in the limit $\alpha \to +\infty$. But one is faced with a first surprise for $\alpha = 2$, where a flat temperature profile is restored, although,
as numerics clearly indicates, the mechanism of
transport along the chain certainly dominates over the parallel transport process. In the light of what discussed in the
Section \ref{sec:chim} this appear as a possible manifestation of a chimera ballistic regime, although there is no
simple argument allowing us to invoke a relation of this special case with an integrable approximation, if any. 
The fact that the case $\alpha=2$ is characterized by a somehow "weaker nonintegrability" is confirmed also for a related model \cite{Bagchi2017,wang2019extremely}. This can be traced back to 
the fact that in this case the lattice supports a special type of free-tail  
localized excitations (travelling discrete breathers) that
enhance energy transfer \cite{wang2019extremely}.  

A complementary approach is the analysis of space-time scaling of 
equilibrium correlations, that in the short-range case yields useful
information by the dynamical exponent $z$ \cite{Spohn2014}.
A numerical study of  the
structure factors of the FPUT model \cite{DiCintio2019},
shows that for $\alpha > 1$ the dynamical exponent $z$ depends on $\alpha$ in a way that is certainly
different form the one that could be expected on the basis of the theory of L\'evy processes. 
Moreover, upon adding the cubic term $\frac{1}{3}(q_i-q_j)^3$ to the potential 
one recovers the same dependence of $z$ on $\alpha$, up to $\alpha = 5$. This is again a surprise, because in the limit
$\alpha \to +\infty$ the cubic and quartic versions of the FPUT-model should converge to different values of $z$.
It is a matter of fact that no theoretical explanation is, at present, available to cope with this challenging scenario: in particular
no hydrodynamic description is, to the best of our knowledge, available.

\subsection{Anomalous transport via the Multiparticle Collision Method}
\label{sec:mpc}

So far we have discussed the case of lattice models. To test the generality of the results
and their universality, it is important to consider more general low-dimensional 
many-body systems like interacting fluids or even plasmas.  
Although molecular dynamics would be the natural choice, it is computationally convenient
to consider effective stochastic processes capable to 
mimic particle interaction through random collisions. A prominent example is the Multi-Particle-Collision (MPC) simulation scheme \cite{Kapral1999}. 
which has been proposed to simulate the mesoscopic dynamics of 
polymers in solution, as well as colloidal and complex fluids. 
Another application is the modeling parallel heat transport in 
edge tokamak plasma \cite{Ciraolo2018}. 
Indeed, in the regimes of interest of magnetic fusion devices, large temperature gradients will build up along the field line joining 
the hot plasma region (hot source), and the colder one close to 
the wall (that act as a sink). Besides this motivation, we 
mention here some results that are pertinent to the problem of anomalous transport.

In brief, the MPC method consists in partitioning the set of $N_p$ point- particles into $N_c$ disjoint cells. 
Within each cell, the local center of mass coordinates and velocity are computed and a rotation of particle velocities around a random axis in the cell's center of mass frame is performed. The rotation angles are fixed
by imposing  that the conserved quantities (energy, momentum) are locally preserved. 
Then, all particles are then propagated freely. Physical details of the interaction can be easily included phenomenologically, for 
instance introducing energy-dependent collision rates \cite{DiCintio2017}. 
Interaction with external reservoirs can be also included by imposing Maxwellian distributions of velocity and chemical potentials on the thermostatted cells \cite{Benenti2014}.  

For the case of a one-dimensional MPC fluid, since the conservation laws are the same as say, the FPUT model, we expect it to belong to the same KPZ universality class of 
anomalous transport. Indeed, numerical measurements of dynamical 
scaling fully confirm this prediction \cite{DiCintio2015}. 
This result is rather robust. The same type of anomalies have been shown to hold also 
for quasi-one-dimensional MPC dynamics, namely in the case of a fluid confined in a box with a large enough aspect ratio \cite{DiCintio2017}.

%\subsection{What's about local equilibrium?}
 
 \subsection{Anomalous heat transport in material science}
 \label{sec:mat}
 The impact of the discovery of anomalous heat transport in anharmonic chains has 
 triggered the search for such an important physical effect in real low-dimensional materials.
 There is nowadays a vast literature, where these kind of phenomena has been predicted
 and experimentally observed and part of this growing research field has been 
 sketched in \cite{Lepri2016}. Here we just want to illustrate two recent contributions 
 which allow the reader to appreciate the relevance of this phenomenon for 
 nanowires  and polymers. We want to point out that an overview on the recent literature in both fields
 is contained in the bibliography of these recent contributions.
 
 In \cite{Upadhyaya2016} the problem of the lattice thermal conductivity
 in Si-Ge nanowires has been tackled by solving the Boltzmann transport equation.
 More precisely, the authors used a Monte Carlo algorithm for sampling the phonon
 mean free path, combined with phenomenological ingredients concerning a suitable
 representation of realistic boundary conditions. It is quite remarkable that they
 find evidence of a heat transport mechanism ruled by a L\'evy walk dynamics 
 of phonon flights through the lattice structure. In particular, the phonon mean free paths
 are found to be characterized by a heavy-tailed distribution, which is 
 associated to an anomalous diffusive behaviour, characterized by
 a size-divergent heat conductivity $\kappa(L) \sim L^{0.33}$. This behavior
 has been checked for system sizes in the range $10 nm < L < 10 \mu m$.
  Note that the phonon mean free path is orders of magnitude smaller
 than this size range.
 It is important to point out that this scenario is robust for different alloy compositions,
 where  the Ge component varies in the range $[6\%, 86\%]$. All of these results
 fully agree with the theoretical expectation that anharmonic chains with leading
 cubic nonlinearity should exhibit a divergent heat conductivity with an 
 exponent $\gamma = \frac{1}{3}$.
 
In \cite{Colombo2018} atomistic simulations have been performed for 
poly(3,4-ethylenedioxythiophene), termed PEDOT, a conjugated polymer
which is considered to be of interest in view of its tunable 
and large electrical conductivity, transparency and air stability \cite{Sun2015}.
The authors simulated this polymer model in $d=1$ and in $d=3$, both 
in equilibrium and nonequilibrium setups. More precisely, equilibrium measurements have been
performed by estimating the dependence of the heat conductivity $\kappa$ on the
system size $L$ by the Green-Kubo formula, where one has to estimate the asymptotic
decay in time of the correlator
of the total energy current (\ref{longtail}). 
The outcome of this analysis has been compared
with the numerics obtained in nonequilibrium conditions.  
The setup adopted in this case is based on a transient measurement of the
effective heat diffusivity $\bar\kappa$. 
%that appears in the Fourier heat equation
%$$ \frac{\partial T}{\partial x} = \bar\kappa
%\frac{\partial^2 T}{\partial x^2}$$
The two halves of the system are initially prepared in two thermalized states  at different temperatures
$T_1$ and $T_2$. By running molecular dynamics simulation one can
measure $\bar\kappa$  as a function of $L$ during the
transient evolution to thermal equilibrium state at temperature $(T_1 + T_2)/2$.
More precisely, the estimate of $\bar\kappa$ relies on the fit of the
time-dependent temperature difference in the two regions (for details see 
Eq.(9) in \cite{Colombo2018}). Finally, the thermal conductivity is obtained by the 
formula $\kappa = \bar\kappa \, \rho\, c_V$
where $\rho$ is the polymer mass density and $c_V$ its specific heat.
The authors have obtained consistent results showing that for the polymer chain 
anomalous diffusion is obeserved, $\kappa(L) \sim L^{\gamma}$ with $\gamma \simeq \frac{1}{2}$,
while for the polymer crystal standard diffusion, i.e. a size-independent finite thermal conductivity $\kappa$,
is recovered. These results are quite remarkable, because they provide a very clear 
confirmation of the role played by space dimension in determining anomalous transport effects. On the other 
hand the authors point out that the exponent $\gamma \simeq \frac{1}{2}$ does not agree with
the expected theoretical value $\frac{1}{3}$, since the phenomenological AMBER nonlinear potential
adopted for the PEDOT model is certainly asymmetric, i.e. dominated by a leading cubic nonlinearity.
Simulations of the polymer chain have been performed for quite large system sizes, namely
$0,376 \mu m < L < 7.526 \mu m$. This notwithstanding, 
we cannot exclude that, as discussed in the previous section, the combination of finite size effects and nonlinearity 
might be at work also in this case, yielding a power-law divergence of $\kappa (L)$ that should be compatible
with a symmetric phenomenological potential. 
%%Citerei qualche lavoro su FPU \beta 

These findings should be compared with those obtained for simpler models: for instance,  
in the case mentioned in Sec. \ref{sec:magn}, the exponent $\gamma = \frac{1}{2}$ has 
been found in a chain model with a quadratic interaction
potential between the beads \cite{Tamaki2017}
%Shuji Tamaki, Makiko Sasada, and Keiji Saito, PRL 119, 110602 (2017)
(notice that the possibility of displacements in both the horizontal and the vertical directions
make this model non--integrable, at variance with the harmonic chain). 
On the other hand, a three-dimensional anharmonic chain with cubic and quartic 
interaction has been shown to belong instead to the KPZ class with $\gamma=1/3$ \cite{Barreto2019}.
We could speculate that in any polymer model
the basic scaling associated to anomalous transport is determined just by the quadratic term, but this is certainly
a conjecture that would deserve further theoretical and numerical investigations.

To conclude, we briefly mention some experimental investigations. Thermal 
properties of nanosized objects have an intrinsic technological interest for 
nanoscale thermal management. In this general context, nanowires
and single-walled nanotubes have been analyzed to look for deviation from the standard Fourier's
law \cite{Chang2016}. Some experimental evidence of anomalous transport in very long
carbon nanotubes has been reported \cite{Lee2017} although the results appear
controversial \cite{Li2017comment}. Experiments demonstrating a non-trivial
length dependence of thermal conductance for molecular chains 
have also been reported \cite{Meier2014}. 
 Admittedly, such experimental 
evidences of anomalous transport are rather limited, and
not exempt from criticism. Heat transfer measurements on nano-sized
objects are notoriously difficult but may possibly be undertaken
in the future, possibly guided by the theoretical insight here
summarized.

\section{Coupled transport}

The purpose of this section is to discuss the relevance of anomalous
diffusion in coupled transport. In particular, we shall focus on 
steady-state transport and use for concreteness the language 
of thermoelectricty \cite{Benenti2016,Benenti2017}, 
where the coupled flows are charge and heat flow (other examples where
the flow coupled to heat is particle or magnetization flow
could be discussed similarly).
Moreover, we shall limit our discussion to power production, 
even though many of the results and open problems 
highlighted below could be readily extended to
refrigeration.
Thermoelectricity is a \emph{steady state} heat engine. Relevant quantities 
to characterize the performance of a generic heat engine, operating 
between a hot reservoir at temperature $T_h$ and a cold 
one at $T_c$, are:
\begin{itemize}
\item
The \emph{efficiency} $\eta=W/Q_h$, where $W$ is the output work and 
$Q_h$ the heat extracted from the hot reservoir.
For cyclic as well as for steady-state heat engines, 
the Carnot efficiency $\eta_C=1-T_c/T_h$ is an upper bound 
for the efficiency $\eta$.
\item
The \emph{output power} $P$. It is common belief that an engine
reaching the Carnot efficiency would require a quasi-static 
transformation, i.e. infinite cycle-time, implying vanishing
power. For steady-state engines this argument is replaced by the 
one that finite currents would imply dissipation, thus forbidding
Carnot efficiency for non-zero power. Hence, it is 
important to consider the power-efficiency trade-off.
This is a key problem in the field of finite-time 
thermodynamics \cite{Andresen2011}, 
in relation to the fundamental thermodynamic
bounds on the performance of heat engines as well as 
to the practical purpose of designing engines that, for a given output power,
work at the maximum possible efficiency.
For classical cyclic heat engines, whose interactions with heat bath 
can be described by a Markov process, it was proved \cite{Saito2016} that
the mean power $P$ has an upper bound
\begin{equation}
P\le \frac{A}{T_c}\,\eta (\eta_c-\eta),
\label{eq:boundSaito}
\end{equation}
where $A$ is a system-specific prefactor
[see also \cite{Seifert2015}
for an analogous linear-response result within the framework
of stochastic thermodynamics \cite{Seifert2012}]. 
While at first sight this bound implies that 
$P\to 0$ when $\eta\to\eta_c$ (and of course when $\eta\to 0$),
so that an engine of that kind with finite power
never attains the Carnot efficiency, one cannot exclude
that the amplitude  $A$ diverges as the efficiency approaches the 
Carnot value \cite{Campisi2016}.

\item
The \emph{fluctuations} in the power output
around its mean value $P$. Indeed, large fluctuations
render heat engines unreliable. Especially for heat engines at the nanoscale,
one expects power fluctuations due, e.g., to thermal noise, which are not 
negligible in comparison with the mean output power. In general, one 
would like to obtain high efficiency (as close as possible to the 
Carnot efficiency), large power, and small fluctuation.
However, a trade-off between these three quantities has 
been proved \cite{Seifert2018} for a broad 
class of steady-state heat engines
(including machines described by suitable rate equations 
or modeled with an overdamped Langevin dynamics):
\begin{equation}
P\,\frac{\eta}{\eta_C-\eta}\frac{T_c}{\Delta_P}\le \frac{1}{2},
\label{eq:seifertbound}
\end{equation}
 where the (steady-state) power fluctuations are given by
\begin{equation}
\Delta_P\equiv \lim_{t\to\infty}[P(t)-P]^2\,t,
\label{eq:TUR}
\end{equation}
where $P(t)$ is the mean delivered power up to time $t$.
Since $P(t)$ converges for $t\to\infty$ to $P$ as $1/\sqrt{t}$, 
an additional factor of $t$ in (\ref{eq:TUR}) is needed to obtain a finite limit for 
$\Delta_P$.
Eq. (\ref{eq:seifertbound}) tells us that efficiency close to Carnot
and high power entail large fluctuations.
We note that the bound (\ref{eq:TUR}) has been recently generalized to 
 periodically driven systems \cite{Seifert2019}.
%A bound relating power, efficiency, and fluctuations has also been
%recently derived \cite{Seifert2019} for periodically driven systems
%described by a Markovian dynamics on a discrete set of states. 
\end{itemize}

\subsection{Linear response}

Within linear response, the relationship between currents and generalized forces is linear \cite{callen,mazur}. In particular, for thermoelectric transport we have
\begin{eqnarray}
\left\{
\begin{array}{l}
j_e=L_{ee} \mathcal{F}_e + L_{eu} \mathcal{F}_h,
\\
\\
j_u=L_{ue} \mathcal{F}_e + L_{uu} \mathcal{F}_h,
\end{array}
\right.
\label{eq:coupledlinear}
\end{eqnarray}
where $j_e$ is the electric current density, $j_u$ is 
the energy current density, and the conjugated generalized forces are 
$\mathcal{F}_e=-\nabla (\mu/eT)$ and 
$\mathcal{F}_h=\nabla(1/T)$, where $\mu$ is the electrochemical potential and $e$ is the electron charge. The coefficients $L_{ab}$ ($a,b=e,u$) are known as kinetic or Onsager coefficients;
we will denote ${\bm L}$ the Onsager matrix with matrix elements $L_{ab}$.
Note that the (total) energy current
$j_u=j_h+(\mu/e) j_e$ is the sum of the heat current $j_h$ and the
electrochemical potential energy current $(\mu/e) j_e$.

The Onsager coefficients must satisfy two fundamental constraints. First, the second law of thermodynamics, that is, the positivity of the entropy production rate, $\dot{s}=\mathcal{F}_e j_e + \mathcal{F}_u j_u\ge 0$, implies
\begin{equation}
L_{ee},
L_{uu}\ge 0,
\quad
L_{ee}L_{uu}
-\frac{1}{4}\,(L_{eu}+L_{ue})^2 \ge 0.
\end{equation}
Second, for systems with time-reversal symmetry, Onsager derived fundamental relations, known as Onsager reciprocal relations: $L_{eu}=L_{ue}$.

The kinetic coefficients $L_{ab}$ are related to the familiar thermoelectric transport coefficients: the electrical conductivity $\sigma$, the thermal conductivity $\kappa$, the thermopower (or Seebeck coefficient) $S$, and the Peltier coefficient $\Pi$:
\begin{equation}
\sigma=-e\,\left(\frac{j_e}{\nabla\mu}\right)_{\nabla T=0}=\frac{L_{ee}}{T},
\label{eq:el_conductivity}
\end{equation}
\begin{equation}
\kappa=-\left(\frac{j_h}{\nabla T}\right)_{j_e=0}=
\frac{1}{T^2}\frac{\det {\bm L}}{L_{ee}},
\label{eq:th_conductivity}
\end{equation}
\begin{equation}
S=-\frac{1}{e}\left(\frac{\nabla \mu}{\nabla T}\right)_{j_e=0}=
\frac{1}{T}\left(\frac{L_{eu}}{L_{ee}}-\frac{\mu}{e}\right),
\label{eq:seebeck}
\end{equation}
\begin{equation}
\Pi=\left(\frac{j_h}{j_e}\right)_{\nabla T=0}
=\frac{L_{ue}}{L_{ee}}-\frac{\mu}{e}.
\label{eq:peltier}
\end{equation}
For systems with time reversal symmetry,  
due to the Onsager reciprocal relations $\Pi=TS$.

The thermoelectric performance is governed by the thermoelectric figure of merit
\begin{equation}
ZT=\frac{\sigma S^2}{\kappa}.
\label{eq:ZT-def}
\end{equation}
Thermodynamics imposes a lower bound on the figure of merit: $ZT\ge 0$. 
Moreover, the thermoelectric conversion efficiency is a monotonous increasing function of $ZT$, 
with $\eta=0$ at $ZT=0$ 
and $\eta\to\eta_C$ in the limit $ZT\to\infty$.
Nowadays, most efficient thermoelectric devices operate at around 
$ZT \approx 1$. On the other hand, it is generally accepted that $ZT > 3-5$ 
is the target value for efficient, commercially competing, 
thermoelectric technology. It is a great challenge to 
increase the thermoelectric efficiency,
since the transport coefficients $S,\sigma,\kappa$ are generally interdependent.
For instance in metals
$\sigma$ and $\kappa$ are proportional according to the Wiedemann-Franz law,
and the thermopower is small, and these facts make
metals poor thermoelectric materials. It is therefore of great importance to
understand tha physical mechanisms that might allow to independently control 
the above transport coefficients.

\subsection{Anomalous transport and efficiency}

The theoretical discussion of the role of 
anomalous (thermal) diffusion in thermoelectric transport 
is based on the Green-Kubo formula.
Such a formula 
%is rooted in the
%fluctuation-dissipation theorem,
%and 
expresses the Onsager coefficients 
in terms of dynamic correlation functions of the corresponding 
currents, computed at thermodynamic equilibrium.
When the current-current correlations 
$\langle j_a(0)j_b(t) \rangle$ ($\langle \, . \,\rangle$ denotes
the canonical average at a given temperature $T$) do not decay
after time averaging, then by definition 
the corresponding Drude weight  
\begin{equation}
D_{ab}=\lim_{t\to\infty}\lim_{\Lambda\to\infty} \frac{1}{2 \Omega(\Lambda) t}
\int_0^t dt \langle j_a(0)j_b(t) \rangle
\end{equation}
is different from zero. Here $\Omega$ is the system's volume and $\Lambda$ 
the system's size along the direction of the currents.
It has been shown \cite{Zotos1997,Zotos2004,Garst2001,Heidrichmeisner2005}
that a non-zero Drude weight $D_{ab}$ is a signature 
of ballistic transport, namely in the thermodynamic limit the corresponding
kinetic coefficient $L_{ab}$ diverges linearly with the system size.
Non-zero Drude weights can be related to 
the existence of relevant conserved quantities,
 which determine a lower bound to 
$D_{ab}$ \cite{Mazur1969,Suzuki1971}.
By definition, a constant of motion $Q$ is relevant
if it is not orthogonal to the currents under consideration, in
thermoelectricity $\langle j_e Q\rangle\ne 0$ and
$\langle j_u Q\rangle\ne 0$.

With regard to thermoelectric efficiency, a theoretical argument
\cite{Benenti2013} predicts that, for systems with a single relevant 
conserved quantity, as it is the case for non-integrable systems
with elastic collisions (momentum-conserving systems), 
the figure of merit $ZT$ 
diverges at the thermodynamics limit, so that the Carnot efficiency
is attained in that limit. 
Indeed, for systems where total momentum along the direction of the currents%%%%%%%%%%%%%%%%%
\begin{figure}
\includegraphics[width=10cm]{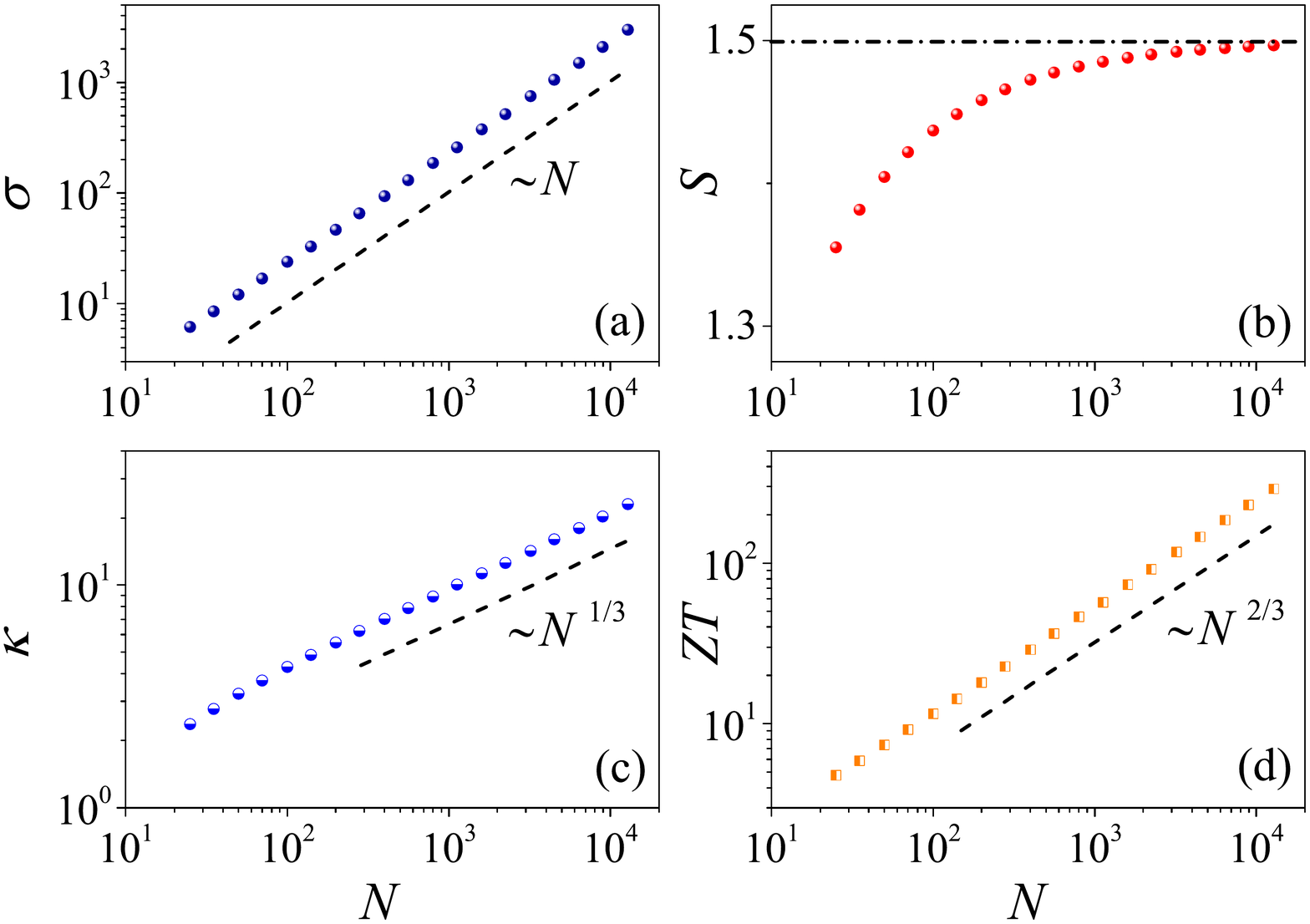}
\caption{Electrical conductivity $\sigma$ (a), Seebeck coefficient $S$ (b), 
thermal conductivity $\kappa$ (c) and 
thermoelectric figure of merit $ZT$ (d) as a function of the mean
number $N$ of particles inside the system, for the one-dimensional, diatomic
hard-point gas. Parameter values: masses $m=1$ and $M=3$,
$T=1$, and $\mu=0$, $k_B=e=1$, system lenhgth $\Lambda$ equal to the number of 
particles. The figure is adapted with permission 
from \cite{Luo2018} (Copyright (2018) by American Physical Society).}
\label{fig:ZT}
\end{figure}
%%%%%%%%%%%%%%%%%
is the only relevant constant of motion, as a consequence of 
ballistic transport  the Onsager coefficients 
$L_{ab}\propto\Lambda$. 
Therefore, the electrical current is ballistic,
$\sigma\sim L_{ee}\sim\Lambda$, while the thermopower is 
asymptotically size-independent, $S\sim L_{eu}/L_{ee}\sim\Lambda^0$.
On the other hand, for such systems
the ballistic contribution to $\det {\bm L}$ is expected to vanish
\cite{Benenti2013}.
Hence 
%$\det {\bm L}$ grows slower than $\Lambda^2$ and therefore 
the thermal conductivity $\kappa\sim\det {\bm L}/L_{ee}$ 
grows sub-ballistically,
$\kappa\sim\Lambda^\gamma$, with $\gamma<1$.
Since $\sigma\sim \Lambda$ and $S\sim\Lambda^0$, we can conclude 
that $ZT\sim\Lambda^{1-\gamma}$. That is,
$ZT$ diverges in the thermodynamic limit $\Lambda\to\infty$.

This result has been illustrated in several models:
in a diatomic chain of hard-point colliding particles 
\cite{Benenti2013} (see Fig.~\ref{fig:ZT}), 
in a two-dimensional system \cite{Benenti2014}, with the dynamics
simulated by the multiparticle collision dynamics method 
mentioned in Section \ref{sec:mpc}
\cite{Kapral1999}, and in a one-dimensional gas of particles with 
screened (nearest neighbour) Coulomb interaction \cite{Chen2015}. 
In all these (classical) models, collisions are elastic and the 
only relevant constant of motion is the 
component of momentum along the direction of the charge and heat flows. 
In the numerical simulations, openings connect the 
system with two electrochemical reservoirs.
The left ($L$) and right ($R$) reservoirs are modeled as ideal gases,
at temperature $T_\gamma$ and electrochemical potential $\mu_\gamma$
($\gamma=L,R$).
A stochastic model of the reservoirs \cite{Mejia2001,Larralde03} is  used:
whenever a particle of the system crosses the opening, which separates
the system from the left or right reservoir, it is removed.
Particles are injected into the system through the openings, with rates
and energy distribution determined by temperature and
electrochemical potential (see, \emph{e.g.}, \cite{Benenti2017}).

%%%%%%%%%%%%%%%%%%
%\begin{figure}
%\includegraphics[width=8.5cm]{ZT.pdf}
%\caption{Electrical conductivity $\sigma$ (a), thermopower $S$ (b), thermal
%conductivity $\kappa$ (c) and figure of merit $ZT$ (d) as a function of the mean
%number $N$ of particles inside the system, for the one-dimensional, diatomic
%hard-point gas. Data are obtained for masses $m=1$, $M=3$,
%$T=1$, and $\mu=0$, $k_B=e=1$, system lenhgth $L$ equal to the number of 
%particles. 
%The figure is adapted with permission 
%from \cite{Luo2018} (Copyright (2018) by American Physical Society).}
%\label{fig:ZT}
%\end{figure}
%%%%%%%%%%%%%%%%%%

We now show that 
systems with anomalous (thermal) diffusion allow a much better power-efficiency
trade-off than achievable by means of non-interacting systems or more 
generally by any system that can be described by the scattering theory.

We first briefly discuss noninteracting systems.
In this case, we can write the charge current 
following the 
Landauer-B\"uttiker scattering theory 
\cite{datta}, adapted to classical physics:
\begin{equation}
j_e=\frac{e}{h}\int_0^\infty d\epsilon\,
[f_L(\epsilon)-f_R(\epsilon)]
{\cal T}(\epsilon),
\end{equation}
where
$f_\gamma(\epsilon)=e^{-\beta_\gamma (\epsilon-\mu_\gamma)}$ is the Maxwell-Boltzmann
distribution function for reservoir $\gamma$ 
and ${\cal T}(\epsilon)$ is the \emph{transmission probability} for
a particle with energy $\epsilon$ to 
go from one reservoir to the other:
$0\le {\cal T}(\epsilon)\le 1$.
Similarly, we obtain the heat current from reservoir $\gamma$ as
\begin{equation}
J_{h,\gamma}=\frac{1}{h}\int_0^\infty d\epsilon\, (\epsilon-\mu_\gamma)
[f_L(\epsilon)-f_R(\epsilon)]
{\cal T}(\epsilon).
\end{equation}
For a given output power $P=(\Delta \mu/e) J_e$
($\Delta\mu=\mu_R-\mu_L>0$ and we set for the sake of simplicity $\mu_L=0$),
the transmission function that maximizes the efficiency
of the heat engine, $\eta(P)=P/J_{h,L}$
($P,J_{h,L}>0$, $T_L>T_R$) was determined in \cite{Luo2018}, closely following 
the method developed for the quantum case in \cite{Whitney2014,Whitney2015}.
The optimal transmission  function
is a boxcar function, ${\cal T} (\epsilon)=1$ 
for $\epsilon_0<\epsilon<\epsilon_1$ and
${\cal T}(\epsilon)=0$ otherwise. 
Here $\epsilon_0=\Delta\mu/\eta_C$ is obtained from the
condition $f_L(\epsilon_0)=f_R(\epsilon_0)$,
 and corresponds to the  special value of energy for which
the flow of particles from left to right is the same as the 
flow from right to left. 
Thus, if particles only flow at energy $\epsilon_0$, the flow 
can be considered as ``reversible'', in a thermodynamic sense.
The energy $\epsilon_1$ and
$\Delta \mu$ are determined numerically in the optimization
procedure \cite{Whitney2014,Whitney2015,Luo2018}.
The maximum achievable
power is obtained when $\epsilon_1\to\infty$:
\begin{equation}
P_{\rm max}^{({\rm st})}=A\frac{\pi^2}{h}\,k_B^2\,(\Delta T)^2,
\label{eq:pendry}
\end{equation}
where $\Delta T=T_L-T_R$, $A\approx 0.0373$,
and the superscript st reminds us that the obtained results are  within 
the scattering theory approach.
At small output power, $P/P_{\rm max}^{({\rm st})}\ll 1$,
\be
\eta(P)\le \eta_{\rm max}^{({\rm st})}(P)=\eta_C\left(
1-B\sqrt{\frac{T_R}{T_L}\frac{P}{P_{\rm max}^{({\rm st})}}}
\right),
\label{eq:whitney}
\ee
where $B\approx 0.493$. 
Note that in the limit $P\to 0$ 
the upper bound on efficiency achieves the Carnot efficiency
and the energy window for transmission,
$\delta=\epsilon_1-\epsilon_0\to 0$.
That is, we recover the celebrated delta-energy filtering
mechanism for Carnot efficiency
\cite{Mahan1996, Humphrey2002, Humphrey2005}. 
Hence, the Carnot limit corresponds to the above mentioned
reversible, zero power flow of particles. 

\begin{figure}
\includegraphics[width=8.5cm]{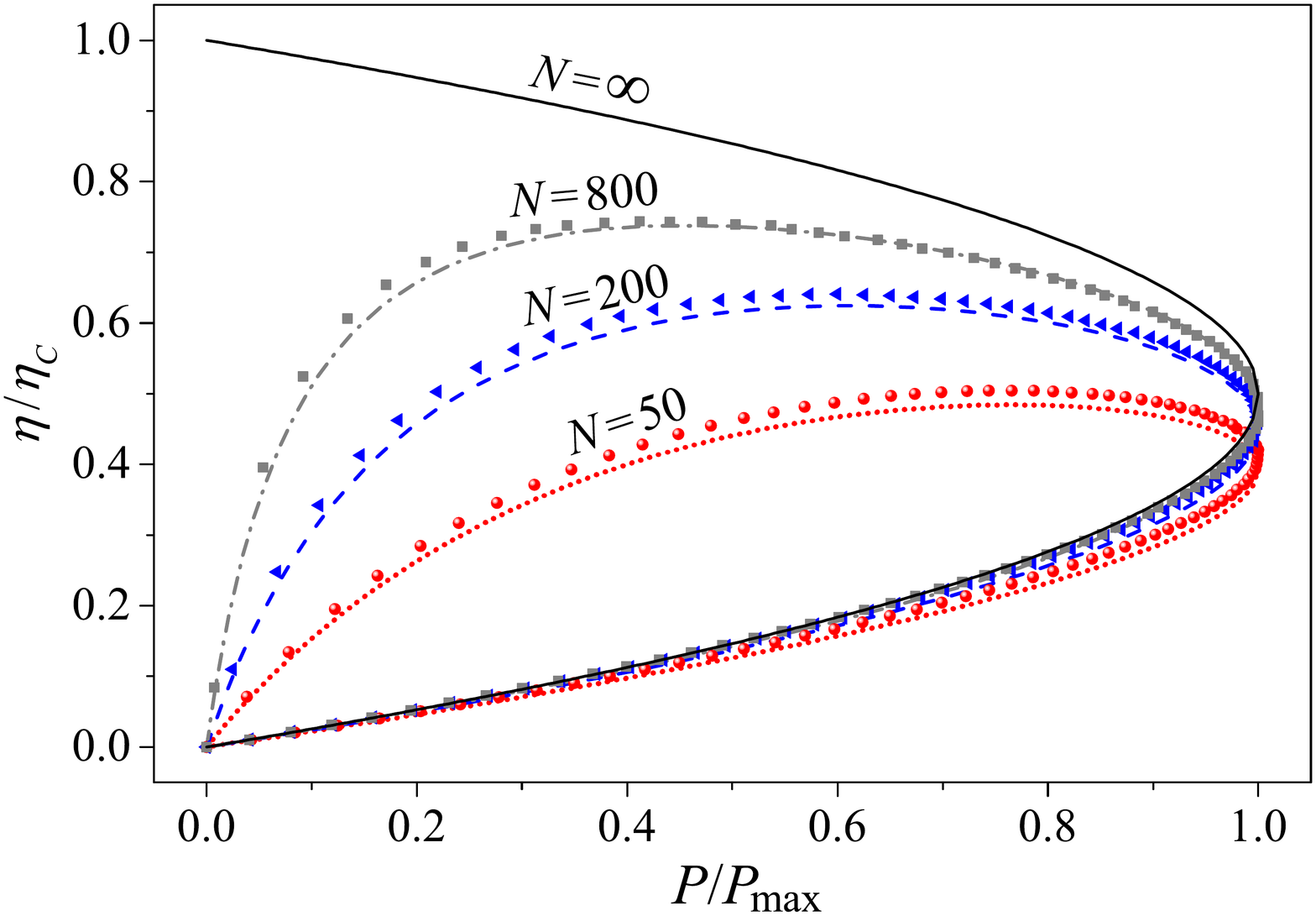}
\caption{Relative efficiency $\eta/\eta_C$ versus normalized power $P/P_{\rm max}$, at different systems sizes.  
The dotted,
dashed, and dot-dashed curves show the linear response predictions, 
at the $ZT(N)$ value corresponding to the given system size 
$\Lambda=N$. The solid line is
the linear response result for $ZT=\infty$ (i.e., $N=\infty$). 
Model an parameter values as Fig.~\ref{fig:ZT},
with $\Delta T=0.2$ ($T_L=1.1$, $T_R=0.9$). Adapted with permission from \cite{Luo2018}  Copyright (2019) by American Physical Society.}
\label{fig:etaP}
\end{figure}

It is clear that selecting transmission
over a small energy window reduces power production. 
We can  then 
expect that a different mechanism to reach Carnot 
efficiency might be more favorable for power production. 
Such expectations are borne out by numerical data for 
the above described interacting
momentum-conserving systems.
For these systems, the Carnot efficiency can be 
reached without delta-energy filtering \cite{Saito2010}, and
the power-efficiency trade-off can be improved.
In Fig.~\ref{fig:etaP}, we show, for a given $\Delta T$ and different system
sizes, $\eta/\eta_C$ as a function of 
$P/P_{\rm max}$. These curves have two branches. Indeed, they are obtained by
increasing $\Delta \mu$ from zero, where trivially $P=0$, 
up to the stopping value, where again
$P=0$.  
In the latter case, power  vanishes because 
the electrochemical potential difference 
becomes too high to be overcome by the temperature difference. 
Power first increases with $\Delta\mu$, up to its maximum value $P=P_{\rm max}$,
and then decreases, leading to a two-branch curve.
Note that, despite the relatively high value of $\Delta T/T=0.2$, 
the numerical results are in rather
good agreement with the universal linear
curves, which only depends on the figure of merit $ZT$ ~\cite{Luo2018}.
Not surprisingly, such agreement improves with
increasing the system size, since $|\nabla T|=\Delta T/N$ decreases when $N$
increases. 
In Fig.~\ref{fig:etaP}, we also show the limiting curve corresponding to
$ZT=\infty$, obtained in momentum-conserving models in the thermodynamic limit
$N\to\infty$. The upper branch of this curve 
is the universal linear response upper bound to
efficiency for a given power $P$. For $P/P_{\rm max}\ll 1$, 
this bound  reads as follows:
\begin{equation}
\eta_{\rm lr}(P)=\eta_C\left(1-\frac{1}{4}\,\frac{P}{P_{\rm max}}
\right),
\label{eq:etalr}
\end{equation}
which is a much less restrictive bound than bound (\ref{eq:whitney}), 
obtained above from scattering theory.
Note that, using the linear response result $P_{\rm max}\propto (\Delta T)^2$,
from (\ref{eq:etalr}) we obtain $P\propto \Delta T\,(\eta_C-\eta)$.
Accordingly, when $\eta\approx \eta_C \propto \Delta T$, we find the same dependence 
as in bound (\ref{eq:boundSaito}), that was obtained in a 
rather different context.

%%%%%%%%%%%%%%%%%
%%%%%%%%%%%%%%%%%

\subsection{Open problems}

Several open questions about the role of anomalous transport in coupled transport remain, 
notably:
\begin{itemize}
\item
As discussed in Sec.~\ref{sec:long},
heat transport in the presence of long-range interactions has been recently 
investigated. However, the 
role of the range of interactions on coupled transport, and in 
particular on the power-efficiency trade-off, is unknown. 
\item
While momentum-conserving systems greatly improve the power-efficiency trade-off
with respect to the noninteracting case,
it is not known how they would behave with respect to the bound
(\ref{eq:seifertbound}),
which simultaneously involves efficiency, power, and fluctuations.
 That bound was obtained within the framework of 
stochastic thermodynamics \cite{Seifert2012}, with the transition 
rates between the system's states which obey the local detailed 
balance principle, without precise modeling of the underlying particle-particle
interactions. On the other hand, in the models described in this 
short review paper stochasticity is confined to the baths and the internal 
system's dynamics plays a crucial role. This allowed us to discuss the 
impact of constants of motion and anomalous transport on the efficiency of heat-to-work conversion and could be relevant also in relation to 
the bound (\ref{eq:seifertbound}).
\item 
While the results of this section 
have been corroborated by numerical simulations of several classical one- and two-dimensional
systems, their extension to the quantum case remains as a challenging task for future investigations.
\item
The discussion of this section neglected phonons. However,
besides their fundamental interest, the results presented here 
could be of practical relevance
in very clean systems where the elastic mean free path of the conducting
particles is much longer than the length scale associated with elastic 
particle-particle collisions, for instance in high-mobility
two-dimensional electron gases at very low temperatures. 
Phonon-free thermoelectricity (more precisely, thermodiffusion) has been 
experimentally realized in the 
context of cold atoms, first 
for weakly interacting particles~\cite{Brantut2013}
and more recently in a regime with strong interactions~\cite{Husmann2018}.
In this latter case, a strong violation of the Wiedemann-Franz law
has been observed. 
Such violations 
could not be explained by the Landauer-B\"uttiker scattering theory. 
It would be interesting to investigate whether
in such systems, where a high thermoelectric efficiency has been observed,
the non-interacting bound on efficiency for a given
power could be outperformed. 
%In these systems, possible applications, rather than for power
%production could, however, be related to the refrigeration
%of atomic gases.
\end{itemize}

\section{Overview}

In spite of the significant progress made over the last decades, the study of anomalous heat transport in nonlinear systems still remains a challenging research field.
While this review focused on some promising directions in classical systems, a main avenue for future investigations should undoubtedly be sought in the quantum domain.
At the quantum level, anomalous  heat transport is considerably less understood than in the classical case due to both conceptual and practical difficulties. The same
definition of thermodynamic observables like temperature, heat, and work, or of the concept of local equilibrium, is problematic in nanoscale systems. For instance, in
solid-state nanodevices we can have structures smaller than the length scale over which electrons relax to a local equilibrium due to electron-electron or
electron-phonon interactions. Consequently, quantum interference effects, quantum correlations, and quantum fluctuation effects should be taken into
account~\cite{Benenti2017}.
In particular, many-body localization provides a mechanism for thermalization to fail in strongly disordered systems, with anomalous transport in the vicinity of the
transition between many-body localized and ergodic phases~\cite{Moore2016,Abanin2019}.

From a practical viewpoint, one has to face the computational complexity of simulating many-body open quantum systems, with the size of the Hilbert space growing
exponentially with the number of particles. Notwithstanding these difficulties, a time-dependent density matrix renormalization group method allows the computation of
transport properties of integrable and nonintegrable quantum spin chains driven by local (Lindblad) operators acting close to their boundaries~\cite{Prosen2009}. Sizes
up to $n\sim 100$ spins can be simulated, much larger than the $n\sim 20$ spins achievable with other  methods like Monte Carlo wavefunction
approaches~\cite{Mejia-Monasterio2007}.
The obtained results confirm the relevance of constants of motion on transport properties, with integrable systems that exhibit ballistic heat transport, while for
quantum chaotic systems heat transport is normal (Fourier's law, see ~\cite{Saito2003,Monasterio2005}). In passing, we note that for magnetization transport in some
integrable models like XXZ one can obtain diffusive behavior (Fick's law, see~\cite{Prosen2009}).

From a thermodynamic perspective, the use of local Lindblad operators is problematic. With the exception of quantum chaotic systems, such operators do not drive the
system to a grand-canonical state~\cite{Znidaric2010}.
Furthermore, the use of local Lindblad baths may result in apparent violations of the second law of thermodynamics \cite{Levy2014}.
Global Lindblad dissipators are free from such problems and can be used to simulate heat transport, see e.g. \cite{Balachandran2019}, but are not practical, in that
limited to very small system sizes.
Furthermore, it is crucial for the description of quantum heat engines in the extreme case where the working medium may even consist of a single two-level quantum
system, to take into account medium-reservoir quantum correlations as well as non-Markovian effects, not included in the standard, weak-coupling Lindblad description of
quantum open systems. For first steps in this challenging direction, see \cite{Carrega2015,Plenio2019,Ankerhold2019}. The investigation of anomalous heat transport in
such regime is \textit{terra incognita}.

\section*{Author Contributions}
All Authors contributed equally to the present work.
\section*{Conflict of Interest}
The authors declare that the research was conducted in the absence of any commercial or financial relationships that could be construed as a potential conflict of interest.
\section*{Acknowledgements}
%%%%%%%%%%%%%%%
SL and RL acknowledge partial support from project MIUR-PRIN2017 \textit{Coarse-grained description for non-equilibrium systems and transport phenomena (CO-NEST)} n. 201798CZL and the Eurofusion Enabling Research ENR-MFE19.CEA-06 \textit{Emissive divertor} and thank P. Di Cintio and S. Iubini for 
their unvaluable help.

\bibliography{bibfrontiers}

\begin{thebibliography}{127}
\expandafter\ifx\csname natexlab\endcsname\relax\def\natexlab#1{#1}\fi
\expandafter\ifx\csname bibnamefont\endcsname\relax
  \def\bibnamefont#1{#1}\fi
\expandafter\ifx\csname bibfnamefont\endcsname\relax
  \def\bibfnamefont#1{#1}\fi
\expandafter\ifx\csname citenamefont\endcsname\relax
  \def\citenamefont#1{#1}\fi
\expandafter\ifx\csname url\endcsname\relax
  \def\url#1{\texttt{#1}}\fi
\expandafter\ifx\csname urlprefix\endcsname\relax\def\urlprefix{URL }\fi
\providecommand{\bibinfo}[2]{#2}
\providecommand{\eprint}[2][]{\url{#2}}

\bibitem[{\citenamefont{Klages et~al.}(2008)\citenamefont{Klages, Radons, and
  Sokolov}}]{klages2008anomalous}
\bibinfo{author}{\bibfnamefont{R.}~\bibnamefont{Klages}},
  \bibinfo{author}{\bibfnamefont{G.}~\bibnamefont{Radons}}, \bibnamefont{and}
  \bibinfo{author}{\bibfnamefont{I.~M.} \bibnamefont{Sokolov}},
  \emph{\bibinfo{title}{Anomalous transport: foundations and applications}}
  (\bibinfo{publisher}{John Wiley \& Sons}, \bibinfo{year}{2008}).

\bibitem[{\citenamefont{Benenti et~al.}(2017)\citenamefont{Benenti, Casati,
  Saito, and Whitney}}]{Benenti2017}
\bibinfo{author}{\bibfnamefont{G.}~\bibnamefont{Benenti}},
  \bibinfo{author}{\bibfnamefont{G.}~\bibnamefont{Casati}},
  \bibinfo{author}{\bibfnamefont{K.}~\bibnamefont{Saito}}, \bibnamefont{and}
  \bibinfo{author}{\bibfnamefont{R.~S.} \bibnamefont{Whitney}},
  \bibinfo{journal}{Physics Reports} \textbf{\bibinfo{volume}{694}},
  \bibinfo{pages}{1 } (\bibinfo{year}{2017}), ISSN \bibinfo{issn}{0370-1573},
  \bibinfo{note}{fundamental aspects of steady-state conversion of heat to work
  at the nanoscale},
  \urlprefix\url{http://www.sciencedirect.com/science/article/pii/S0370157317301540}.

\bibitem[{\citenamefont{Lepri et~al.}(1997)\citenamefont{Lepri, Livi, and
  Politi}}]{LLP97}
\bibinfo{author}{\bibfnamefont{S.}~\bibnamefont{Lepri}},
  \bibinfo{author}{\bibfnamefont{R.}~\bibnamefont{Livi}}, \bibnamefont{and}
  \bibinfo{author}{\bibfnamefont{A.}~\bibnamefont{Politi}},
  \bibinfo{journal}{Phys. Rev. Lett.} \textbf{\bibinfo{volume}{78}},
  \bibinfo{pages}{1896} (\bibinfo{year}{1997}), ISSN \bibinfo{issn}{0031-9007}.

\bibitem[{\citenamefont{Lepri et~al.}(1998)\citenamefont{Lepri, Livi, and
  Politi}}]{Lepri98a}
\bibinfo{author}{\bibfnamefont{S.}~\bibnamefont{Lepri}},
  \bibinfo{author}{\bibfnamefont{R.}~\bibnamefont{Livi}}, \bibnamefont{and}
  \bibinfo{author}{\bibfnamefont{A.}~\bibnamefont{Politi}},
  \bibinfo{journal}{Europhys. Lett.} \textbf{\bibinfo{volume}{43}},
  \bibinfo{pages}{271} (\bibinfo{year}{1998}), ISSN \bibinfo{issn}{0295-5075}.

\bibitem[{\citenamefont{Chang}(2016)}]{Chang2016}
\bibinfo{author}{\bibfnamefont{C.-W.} \bibnamefont{Chang}}, in
  \emph{\bibinfo{booktitle}{Thermal Transport in Low Dimensions: From
  Statistical Physics to Nanoscale Heat Transfer}}, edited by
  \bibinfo{editor}{\bibfnamefont{S.}~\bibnamefont{Lepri}}
  (\bibinfo{publisher}{Springer International Publishing},
  \bibinfo{year}{2016}), pp. \bibinfo{pages}{305--338}, ISBN
  \bibinfo{isbn}{978-3-319-29261-8}.

\bibitem[{\citenamefont{Balandin}(2011)}]{Balandin2011}
\bibinfo{author}{\bibfnamefont{A.~A.} \bibnamefont{Balandin}},
  \bibinfo{journal}{Nat. Mater.} \textbf{\bibinfo{volume}{10}},
  \bibinfo{pages}{569} (\bibinfo{year}{2011}).

\bibitem[{\citenamefont{Lepri et~al.}(2003{\natexlab{a}})\citenamefont{Lepri,
  Livi, and Politi}}]{LLP03}
\bibinfo{author}{\bibfnamefont{S.}~\bibnamefont{Lepri}},
  \bibinfo{author}{\bibfnamefont{R.}~\bibnamefont{Livi}}, \bibnamefont{and}
  \bibinfo{author}{\bibfnamefont{A.}~\bibnamefont{Politi}},
  \bibinfo{journal}{Phys. Rep.} \textbf{\bibinfo{volume}{377}},
  \bibinfo{pages}{1} (\bibinfo{year}{2003}{\natexlab{a}}).

\bibitem[{\citenamefont{Dhar}(2008)}]{DHARREV}
\bibinfo{author}{\bibfnamefont{A.}~\bibnamefont{Dhar}}, \bibinfo{journal}{Adv.
  Phys.} \textbf{\bibinfo{volume}{57}}, \bibinfo{pages}{457}
  (\bibinfo{year}{2008}).

\bibitem[{\citenamefont{Lepri}(2016)}]{Lepri2016}
\bibinfo{editor}{\bibfnamefont{S.}~\bibnamefont{Lepri}}, ed.,
  \emph{\bibinfo{title}{Thermal transport in low dimensions: from statistical
  physics to nanoscale heat transfer}}, vol. \bibinfo{volume}{921} of
  \emph{\bibinfo{series}{Lect. Notes Phys}}
  (\bibinfo{publisher}{Springer-Verlag, Berlin Heidelberg},
  \bibinfo{year}{2016}).

\bibitem[{\citenamefont{Dhar et~al.}(2019{\natexlab{a}})\citenamefont{Dhar,
  Kundu, and Kundu}}]{Dhar2019}
\bibinfo{author}{\bibfnamefont{A.}~\bibnamefont{Dhar}},
  \bibinfo{author}{\bibfnamefont{A.}~\bibnamefont{Kundu}}, \bibnamefont{and}
  \bibinfo{author}{\bibfnamefont{A.}~\bibnamefont{Kundu}},
  \bibinfo{journal}{Frontiers in Physics} \textbf{\bibinfo{volume}{7}},
  \bibinfo{pages}{159} (\bibinfo{year}{2019}{\natexlab{a}}),
  \urlprefix\url{https://www.frontiersin.org/article/10.3389/fphy.2019.00159}.

\bibitem[{\citenamefont{Aoki and Kusnezov}(2000)}]{Aoki00}
\bibinfo{author}{\bibfnamefont{K.}~\bibnamefont{Aoki}} \bibnamefont{and}
  \bibinfo{author}{\bibfnamefont{D.}~\bibnamefont{Kusnezov}},
  \bibinfo{journal}{Phys. Lett. A} \textbf{\bibinfo{volume}{265}},
  \bibinfo{pages}{250} (\bibinfo{year}{2000}).

\bibitem[{\citenamefont{Giardin{\'a} et~al.}(2000)\citenamefont{Giardin{\'a},
  Livi, Politi, and Vassalli}}]{Giardina99}
\bibinfo{author}{\bibfnamefont{C.}~\bibnamefont{Giardin{\'a}}},
  \bibinfo{author}{\bibfnamefont{R.}~\bibnamefont{Livi}},
  \bibinfo{author}{\bibfnamefont{A.}~\bibnamefont{Politi}}, \bibnamefont{and}
  \bibinfo{author}{\bibfnamefont{M.}~\bibnamefont{Vassalli}},
  \bibinfo{journal}{Phys. Rev. Lett.} \textbf{\bibinfo{volume}{84}},
  \bibinfo{pages}{2144} (\bibinfo{year}{2000}).

\bibitem[{\citenamefont{Gendelman and Savin}(2000)}]{Gendelman2000}
\bibinfo{author}{\bibfnamefont{O.~V.} \bibnamefont{Gendelman}}
  \bibnamefont{and} \bibinfo{author}{\bibfnamefont{A.~V.} \bibnamefont{Savin}},
  \bibinfo{journal}{Phys. Rev. Lett.} \textbf{\bibinfo{volume}{84}},
  \bibinfo{pages}{2381} (\bibinfo{year}{2000}).

\bibitem[{\citenamefont{Iubini et~al.}(2012)\citenamefont{Iubini, Lepri, and
  Politi}}]{Iubini2012}
\bibinfo{author}{\bibfnamefont{S.}~\bibnamefont{Iubini}},
  \bibinfo{author}{\bibfnamefont{S.}~\bibnamefont{Lepri}}, \bibnamefont{and}
  \bibinfo{author}{\bibfnamefont{A.}~\bibnamefont{Politi}},
  \bibinfo{journal}{Phys. Rev. E} \textbf{\bibinfo{volume}{86}},
  \bibinfo{pages}{011108} (\bibinfo{year}{2012}).

\bibitem[{\citenamefont{Mendl and Spohn}(2015)}]{Mendl2015}
\bibinfo{author}{\bibfnamefont{C.~B.} \bibnamefont{Mendl}} \bibnamefont{and}
  \bibinfo{author}{\bibfnamefont{H.}~\bibnamefont{Spohn}}, \bibinfo{journal}{J.
  Stat. Mech: Theory Exp.} \textbf{\bibinfo{volume}{2015}},
  \bibinfo{pages}{P08028} (\bibinfo{year}{2015}).

\bibitem[{\citenamefont{Lippi and Livi}(2000)}]{Lippi00}
\bibinfo{author}{\bibfnamefont{A.}~\bibnamefont{Lippi}} \bibnamefont{and}
  \bibinfo{author}{\bibfnamefont{R.}~\bibnamefont{Livi}}, \bibinfo{journal}{J.
  Stat. Phys.} \textbf{\bibinfo{volume}{100}}, \bibinfo{pages}{1147}
  (\bibinfo{year}{2000}), ISSN \bibinfo{issn}{0022-4715}.

\bibitem[{\citenamefont{Wang et~al.}(2012)\citenamefont{Wang, Hu, and
  Li}}]{Wang2012}
\bibinfo{author}{\bibfnamefont{L.}~\bibnamefont{Wang}},
  \bibinfo{author}{\bibfnamefont{B.}~\bibnamefont{Hu}}, \bibnamefont{and}
  \bibinfo{author}{\bibfnamefont{B.}~\bibnamefont{Li}}, \bibinfo{journal}{Phys.
  Rev. E} \textbf{\bibinfo{volume}{86}}, \bibinfo{pages}{040101}
  (\bibinfo{year}{2012}).

\bibitem[{\citenamefont{Di~Cintio et~al.}(2017)\citenamefont{Di~Cintio, Livi,
  Lepri, and Ciraolo}}]{DiCintio2017}
\bibinfo{author}{\bibfnamefont{P.}~\bibnamefont{Di~Cintio}},
  \bibinfo{author}{\bibfnamefont{R.}~\bibnamefont{Livi}},
  \bibinfo{author}{\bibfnamefont{S.}~\bibnamefont{Lepri}}, \bibnamefont{and}
  \bibinfo{author}{\bibfnamefont{G.}~\bibnamefont{Ciraolo}},
  \bibinfo{journal}{Phys. Rev. E} \textbf{\bibinfo{volume}{95}},
  \bibinfo{pages}{043203} (\bibinfo{year}{2017}),
  \urlprefix\url{https://link.aps.org/doi/10.1103/PhysRevE.95.043203}.

\bibitem[{\citenamefont{van Beijeren}(2012)}]{VanBeijeren2012}
\bibinfo{author}{\bibfnamefont{H.}~\bibnamefont{van Beijeren}},
  \bibinfo{journal}{Phys. Rev. Lett.} \textbf{\bibinfo{volume}{108}},
  \bibinfo{pages}{180601} (\bibinfo{year}{2012}).

\bibitem[{\citenamefont{Spohn}(2014{\natexlab{a}})}]{Spohn2014}
\bibinfo{author}{\bibfnamefont{H.}~\bibnamefont{Spohn}}, \bibinfo{journal}{J.
  Stat. Phys.} \textbf{\bibinfo{volume}{154}}, \bibinfo{pages}{1191}
  (\bibinfo{year}{2014}{\natexlab{a}}).

\bibitem[{\citenamefont{Popkov et~al.}(2015)\citenamefont{Popkov,
  Schadschneider, Schmidt, and Sch{\"u}tz}}]{Popkov2015}
\bibinfo{author}{\bibfnamefont{V.}~\bibnamefont{Popkov}},
  \bibinfo{author}{\bibfnamefont{A.}~\bibnamefont{Schadschneider}},
  \bibinfo{author}{\bibfnamefont{J.}~\bibnamefont{Schmidt}}, \bibnamefont{and}
  \bibinfo{author}{\bibfnamefont{G.~M.} \bibnamefont{Sch{\"u}tz}},
  \bibinfo{journal}{Proceedings of the National Academy of Sciences}
  \textbf{\bibinfo{volume}{112}}, \bibinfo{pages}{12645}
  (\bibinfo{year}{2015}).

\bibitem[{\citenamefont{Lee-Dadswell et~al.}(2005)\citenamefont{Lee-Dadswell,
  Nickel, and Gray}}]{Lee-Dadswell05}
\bibinfo{author}{\bibfnamefont{G.~R.} \bibnamefont{Lee-Dadswell}},
  \bibinfo{author}{\bibfnamefont{B.~G.} \bibnamefont{Nickel}},
  \bibnamefont{and} \bibinfo{author}{\bibfnamefont{C.~G.} \bibnamefont{Gray}},
  \bibinfo{journal}{Phys. Rev. E} \textbf{\bibinfo{volume}{72}},
  \bibinfo{eid}{031202} (\bibinfo{year}{2005}).

\bibitem[{\citenamefont{Delfini et~al.}(2007)\citenamefont{Delfini, Lepri,
  Livi, and Politi}}]{Delfini07b}
\bibinfo{author}{\bibfnamefont{L.}~\bibnamefont{Delfini}},
  \bibinfo{author}{\bibfnamefont{S.}~\bibnamefont{Lepri}},
  \bibinfo{author}{\bibfnamefont{R.}~\bibnamefont{Livi}}, \bibnamefont{and}
  \bibinfo{author}{\bibfnamefont{A.}~\bibnamefont{Politi}},
  \bibinfo{journal}{J. Stat. Mech.: Theory and Experiment} p.
  \bibinfo{pages}{P02007} (\bibinfo{year}{2007}).

\bibitem[{\citenamefont{Lukkarinen and Spohn}(2008)}]{Lukkarinen2008}
\bibinfo{author}{\bibfnamefont{J.}~\bibnamefont{Lukkarinen}} \bibnamefont{and}
  \bibinfo{author}{\bibfnamefont{H.}~\bibnamefont{Spohn}},
  \bibinfo{journal}{Communications on Pure and Applied Mathematics}
  \textbf{\bibinfo{volume}{61}}, \bibinfo{pages}{1753} (\bibinfo{year}{2008}),
  ISSN \bibinfo{issn}{1097-0312}.

\bibitem[{\citenamefont{Lepri et~al.}(2003{\natexlab{b}})\citenamefont{Lepri,
  Livi, and Politi}}]{Lepri03}
\bibinfo{author}{\bibfnamefont{S.}~\bibnamefont{Lepri}},
  \bibinfo{author}{\bibfnamefont{R.}~\bibnamefont{Livi}}, \bibnamefont{and}
  \bibinfo{author}{\bibfnamefont{A.}~\bibnamefont{Politi}},
  \bibinfo{journal}{Phys. Rev. E} \textbf{\bibinfo{volume}{68}},
  \bibinfo{pages}{067102} (\bibinfo{year}{2003}{\natexlab{b}}), ISSN
  \bibinfo{issn}{1063-651X}.

\bibitem[{\citenamefont{Wang and Wang}(2011)}]{Wang2011}
\bibinfo{author}{\bibfnamefont{L.}~\bibnamefont{Wang}} \bibnamefont{and}
  \bibinfo{author}{\bibfnamefont{T.}~\bibnamefont{Wang}}, \bibinfo{journal}{EPL
  (Europhysics Letters)} \textbf{\bibinfo{volume}{93}}, \bibinfo{pages}{54002}
  (\bibinfo{year}{2011}).

\bibitem[{\citenamefont{Basile et~al.}(2006)\citenamefont{Basile, Bernardin,
  and Olla}}]{BBO06}
\bibinfo{author}{\bibfnamefont{G.}~\bibnamefont{Basile}},
  \bibinfo{author}{\bibfnamefont{C.}~\bibnamefont{Bernardin}},
  \bibnamefont{and} \bibinfo{author}{\bibfnamefont{S.}~\bibnamefont{Olla}},
  \bibinfo{journal}{Phys. Rev. Lett.} \textbf{\bibinfo{volume}{96}},
  \bibinfo{pages}{204303} (\bibinfo{year}{2006}).

\bibitem[{\citenamefont{Lepri et~al.}(2009)\citenamefont{Lepri,
  Mej{\'\i}a-Monasterio, and Politi}}]{Lepri2009}
\bibinfo{author}{\bibfnamefont{S.}~\bibnamefont{Lepri}},
  \bibinfo{author}{\bibfnamefont{C.}~\bibnamefont{Mej{\'\i}a-Monasterio}},
  \bibnamefont{and} \bibinfo{author}{\bibfnamefont{A.}~\bibnamefont{Politi}},
  \bibinfo{journal}{J. Phys. A: Math. Theor.} \textbf{\bibinfo{volume}{42}},
  \bibinfo{pages}{025001} (\bibinfo{year}{2009}).

\bibitem[{\citenamefont{Lee-Dadswell}(2015)}]{Lee-Dadswell2015}
\bibinfo{author}{\bibfnamefont{G.}~\bibnamefont{Lee-Dadswell}},
  \bibinfo{journal}{Phys. Rev. E} \textbf{\bibinfo{volume}{91}},
  \bibinfo{pages}{032102} (\bibinfo{year}{2015}).

\bibitem[{\citenamefont{Zaburdaev et~al.}(2015)\citenamefont{Zaburdaev,
  Denisov, and Klafter}}]{Zaburdaev2015}
\bibinfo{author}{\bibfnamefont{V.}~\bibnamefont{Zaburdaev}},
  \bibinfo{author}{\bibfnamefont{S.}~\bibnamefont{Denisov}}, \bibnamefont{and}
  \bibinfo{author}{\bibfnamefont{J.}~\bibnamefont{Klafter}},
  \bibinfo{journal}{Rev. Mod. Phys.} \textbf{\bibinfo{volume}{87}},
  \bibinfo{pages}{483} (\bibinfo{year}{2015}).

\bibitem[{\citenamefont{Cipriani et~al.}(2005)\citenamefont{Cipriani, Denisov,
  and Politi}}]{Cipriani05}
\bibinfo{author}{\bibfnamefont{P.}~\bibnamefont{Cipriani}},
  \bibinfo{author}{\bibfnamefont{S.}~\bibnamefont{Denisov}}, \bibnamefont{and}
  \bibinfo{author}{\bibfnamefont{A.}~\bibnamefont{Politi}},
  \bibinfo{journal}{Phys. Rev. Lett.} \textbf{\bibinfo{volume}{94}},
  \bibinfo{eid}{244301} (\bibinfo{year}{2005}).

\bibitem[{\citenamefont{Lepri and Politi}(2011)}]{Lepri2011}
\bibinfo{author}{\bibfnamefont{S.}~\bibnamefont{Lepri}} \bibnamefont{and}
  \bibinfo{author}{\bibfnamefont{A.}~\bibnamefont{Politi}},
  \bibinfo{journal}{Phys. Rev. E} \textbf{\bibinfo{volume}{83}},
  \bibinfo{pages}{030107} (\bibinfo{year}{2011}).

\bibitem[{\citenamefont{Dhar et~al.}(2013)\citenamefont{Dhar, Saito, and
  Derrida}}]{Dhar2013}
\bibinfo{author}{\bibfnamefont{A.}~\bibnamefont{Dhar}},
  \bibinfo{author}{\bibfnamefont{K.}~\bibnamefont{Saito}}, \bibnamefont{and}
  \bibinfo{author}{\bibfnamefont{B.}~\bibnamefont{Derrida}},
  \bibinfo{journal}{Phys. Rev. E} \textbf{\bibinfo{volume}{87}},
  \bibinfo{pages}{010103} (\bibinfo{year}{2013}).

\bibitem[{\citenamefont{Denisov et~al.}(2003)\citenamefont{Denisov, Klafter,
  and Urbakh}}]{Denisov03}
\bibinfo{author}{\bibfnamefont{S.}~\bibnamefont{Denisov}},
  \bibinfo{author}{\bibfnamefont{J.}~\bibnamefont{Klafter}}, \bibnamefont{and}
  \bibinfo{author}{\bibfnamefont{M.}~\bibnamefont{Urbakh}},
  \bibinfo{journal}{Phys. Rev. Lett.} \textbf{\bibinfo{volume}{91}},
  \bibinfo{pages}{194301} (\bibinfo{year}{2003}).

\bibitem[{\citenamefont{Liu et~al.}(2014)\citenamefont{Liu, Haenggi, Li, Ren,
  and Li}}]{Liu2014}
\bibinfo{author}{\bibfnamefont{S.}~\bibnamefont{Liu}},
  \bibinfo{author}{\bibfnamefont{P.}~\bibnamefont{Haenggi}},
  \bibinfo{author}{\bibfnamefont{N.}~\bibnamefont{Li}},
  \bibinfo{author}{\bibfnamefont{J.}~\bibnamefont{Ren}}, \bibnamefont{and}
  \bibinfo{author}{\bibfnamefont{B.}~\bibnamefont{Li}}, \bibinfo{journal}{Phys.
  Rev. Lett.} \textbf{\bibinfo{volume}{112}}, \bibinfo{pages}{040601}
  (\bibinfo{year}{2014}).

\bibitem[{\citenamefont{Mendl and Spohn}(2014)}]{Mendl2014}
\bibinfo{author}{\bibfnamefont{C.~B.} \bibnamefont{Mendl}} \bibnamefont{and}
  \bibinfo{author}{\bibfnamefont{H.}~\bibnamefont{Spohn}},
  \bibinfo{journal}{Phys. Rev. E} \textbf{\bibinfo{volume}{90}},
  \bibinfo{pages}{012147} (\bibinfo{year}{2014}).

\bibitem[{\citenamefont{Jara et~al.}(2015)\citenamefont{Jara, Komorowski, and
  Olla}}]{Jara2015}
\bibinfo{author}{\bibfnamefont{M.}~\bibnamefont{Jara}},
  \bibinfo{author}{\bibfnamefont{T.}~\bibnamefont{Komorowski}},
  \bibnamefont{and} \bibinfo{author}{\bibfnamefont{S.}~\bibnamefont{Olla}},
  \bibinfo{journal}{Communications in Mathematical Physics}
  \textbf{\bibinfo{volume}{339}}, \bibinfo{pages}{407} (\bibinfo{year}{2015}),
  ISSN \bibinfo{issn}{1432-0916}.

\bibitem[{\citenamefont{Bernardin et~al.}(2016)\citenamefont{Bernardin,
  Gon{\c{c}}alves, and Jara}}]{Bernardin2016}
\bibinfo{author}{\bibfnamefont{C.}~\bibnamefont{Bernardin}},
  \bibinfo{author}{\bibfnamefont{P.}~\bibnamefont{Gon{\c{c}}alves}},
  \bibnamefont{and} \bibinfo{author}{\bibfnamefont{M.}~\bibnamefont{Jara}},
  \bibinfo{journal}{Archive for Rational Mechanics and Analysis}
  \textbf{\bibinfo{volume}{220}}, \bibinfo{pages}{505} (\bibinfo{year}{2016}).

\bibitem[{\citenamefont{Bernardin and Stoltz}(2012)}]{Bernardin2012}
\bibinfo{author}{\bibfnamefont{C.}~\bibnamefont{Bernardin}} \bibnamefont{and}
  \bibinfo{author}{\bibfnamefont{G.}~\bibnamefont{Stoltz}},
  \bibinfo{journal}{Nonlinearity} \textbf{\bibinfo{volume}{25}},
  \bibinfo{pages}{1099} (\bibinfo{year}{2012}).

\bibitem[{\citenamefont{Bernardin and Gon{\c{c}}alves}(2014)}]{Bernardin2014}
\bibinfo{author}{\bibfnamefont{C.}~\bibnamefont{Bernardin}} \bibnamefont{and}
  \bibinfo{author}{\bibfnamefont{P.}~\bibnamefont{Gon{\c{c}}alves}},
  \bibinfo{journal}{Communications in Mathematical Physics}
  \textbf{\bibinfo{volume}{325}}, \bibinfo{pages}{291} (\bibinfo{year}{2014}).

\bibitem[{\citenamefont{Delfini et~al.}(2010)\citenamefont{Delfini, Lepri,
  Livi, Mejia-Monasterio, and Politi}}]{Delfini10}
\bibinfo{author}{\bibfnamefont{L.}~\bibnamefont{Delfini}},
  \bibinfo{author}{\bibfnamefont{S.}~\bibnamefont{Lepri}},
  \bibinfo{author}{\bibfnamefont{R.}~\bibnamefont{Livi}},
  \bibinfo{author}{\bibfnamefont{C.}~\bibnamefont{Mejia-Monasterio}},
  \bibnamefont{and} \bibinfo{author}{\bibfnamefont{A.}~\bibnamefont{Politi}},
  \bibinfo{journal}{J. Phys. A: Math. Theor.} \textbf{\bibinfo{volume}{43}},
  \bibinfo{pages}{145001} (\bibinfo{year}{2010}).

\bibitem[{\citenamefont{Kundu et~al.}(2019)\citenamefont{Kundu, Bernardin,
  Saito, Kundu, and Dhar}}]{Kundu2019}
\bibinfo{author}{\bibfnamefont{A.}~\bibnamefont{Kundu}},
  \bibinfo{author}{\bibfnamefont{C.}~\bibnamefont{Bernardin}},
  \bibinfo{author}{\bibfnamefont{K.}~\bibnamefont{Saito}},
  \bibinfo{author}{\bibfnamefont{A.}~\bibnamefont{Kundu}}, \bibnamefont{and}
  \bibinfo{author}{\bibfnamefont{A.}~\bibnamefont{Dhar}}, \bibinfo{journal}{J.
  Stat. Mech: Theory Exp.} \textbf{\bibinfo{volume}{2019}},
  \bibinfo{pages}{013205} (\bibinfo{year}{2019}).

\bibitem[{\citenamefont{Iacobucci et~al.}(2010)\citenamefont{Iacobucci, Legoll,
  Olla, and Stoltz}}]{Iacobucci2010}
\bibinfo{author}{\bibfnamefont{A.}~\bibnamefont{Iacobucci}},
  \bibinfo{author}{\bibfnamefont{F.}~\bibnamefont{Legoll}},
  \bibinfo{author}{\bibfnamefont{S.}~\bibnamefont{Olla}}, \bibnamefont{and}
  \bibinfo{author}{\bibfnamefont{G.}~\bibnamefont{Stoltz}},
  \bibinfo{journal}{J. Stat. Phys.} \textbf{\bibinfo{volume}{140}},
  \bibinfo{pages}{336} (\bibinfo{year}{2010}), ISSN \bibinfo{issn}{1572-9613},
  \urlprefix\url{https://doi.org/10.1007/s10955-010-9996-6}.

\bibitem[{\citenamefont{Hurtado and Garrido}(2016)}]{Hurtado2016}
\bibinfo{author}{\bibfnamefont{P.~I.} \bibnamefont{Hurtado}} \bibnamefont{and}
  \bibinfo{author}{\bibfnamefont{P.~L.} \bibnamefont{Garrido}},
  \bibinfo{journal}{Scientific reports} \textbf{\bibinfo{volume}{6}},
  \bibinfo{pages}{38823} (\bibinfo{year}{2016}).

\bibitem[{\citenamefont{Xiong}(2016)}]{Xiong2016}
\bibinfo{author}{\bibfnamefont{D.}~\bibnamefont{Xiong}}, \bibinfo{journal}{J.
  Stat. Mech.: Theory and Experiment} \textbf{\bibinfo{volume}{2016}},
  \bibinfo{pages}{043208} (\bibinfo{year}{2016}).

\bibitem[{\citenamefont{Archana and Barik}(2019)}]{Barik2019}
\bibinfo{author}{\bibfnamefont{G.~R.} \bibnamefont{Archana}} \bibnamefont{and}
  \bibinfo{author}{\bibfnamefont{D.}~\bibnamefont{Barik}},
  \bibinfo{journal}{Phys. Rev. E} \textbf{\bibinfo{volume}{99}},
  \bibinfo{pages}{022103} (\bibinfo{year}{2019}).

\bibitem[{\citenamefont{Basile et~al.}(2007)\citenamefont{Basile, Delfini,
  Lepri, Livi, Olla, and Politi}}]{Basile08}
\bibinfo{author}{\bibfnamefont{G.}~\bibnamefont{Basile}},
  \bibinfo{author}{\bibfnamefont{L.}~\bibnamefont{Delfini}},
  \bibinfo{author}{\bibfnamefont{S.}~\bibnamefont{Lepri}},
  \bibinfo{author}{\bibfnamefont{R.}~\bibnamefont{Livi}},
  \bibinfo{author}{\bibfnamefont{S.}~\bibnamefont{Olla}}, \bibnamefont{and}
  \bibinfo{author}{\bibfnamefont{A.}~\bibnamefont{Politi}},
  \bibinfo{journal}{Eur. Phys J.-Special Topics}
  \textbf{\bibinfo{volume}{151}}, \bibinfo{pages}{85} (\bibinfo{year}{2007}).

\bibitem[{\citenamefont{Zhong et~al.}(2012)\citenamefont{Zhong, Zhang, Wang,
  and Zhao}}]{Zhong2012}
\bibinfo{author}{\bibfnamefont{Y.}~\bibnamefont{Zhong}},
  \bibinfo{author}{\bibfnamefont{Y.}~\bibnamefont{Zhang}},
  \bibinfo{author}{\bibfnamefont{J.}~\bibnamefont{Wang}}, \bibnamefont{and}
  \bibinfo{author}{\bibfnamefont{H.}~\bibnamefont{Zhao}},
  \bibinfo{journal}{Phys. Rev. E} \textbf{\bibinfo{volume}{85}},
  \bibinfo{pages}{060102} (\bibinfo{year}{2012}).

\bibitem[{\citenamefont{Wang et~al.}(2013)\citenamefont{Wang, Hu, and
  Li}}]{Wang2013}
\bibinfo{author}{\bibfnamefont{L.}~\bibnamefont{Wang}},
  \bibinfo{author}{\bibfnamefont{B.}~\bibnamefont{Hu}}, \bibnamefont{and}
  \bibinfo{author}{\bibfnamefont{B.}~\bibnamefont{Li}}, \bibinfo{journal}{Phys.
  Rev. E} \textbf{\bibinfo{volume}{88}}, \bibinfo{pages}{052112}
  (\bibinfo{year}{2013}).

\bibitem[{\citenamefont{Das et~al.}(2014)\citenamefont{Das, Dhar, and
  Narayan}}]{Das2014}
\bibinfo{author}{\bibfnamefont{S.}~\bibnamefont{Das}},
  \bibinfo{author}{\bibfnamefont{A.}~\bibnamefont{Dhar}}, \bibnamefont{and}
  \bibinfo{author}{\bibfnamefont{O.}~\bibnamefont{Narayan}},
  \bibinfo{journal}{J. Stat. Phys.;} \textbf{\bibinfo{volume}{154}},
  \bibinfo{pages}{204} (\bibinfo{year}{2014}), ISSN \bibinfo{issn}{0022-4715}.

\bibitem[{\citenamefont{Chen et~al.}(2014)\citenamefont{Chen, Wang, Casati, and
  Benenti}}]{Chen2014}
\bibinfo{author}{\bibfnamefont{S.}~\bibnamefont{Chen}},
  \bibinfo{author}{\bibfnamefont{J.}~\bibnamefont{Wang}},
  \bibinfo{author}{\bibfnamefont{G.}~\bibnamefont{Casati}}, \bibnamefont{and}
  \bibinfo{author}{\bibfnamefont{G.}~\bibnamefont{Benenti}},
  \bibinfo{journal}{Phys. Rev. E} \textbf{\bibinfo{volume}{90}},
  \bibinfo{pages}{032134} (\bibinfo{year}{2014}),
  \urlprefix\url{https://link.aps.org/doi/10.1103/PhysRevE.90.032134}.

\bibitem[{\citenamefont{Spohn}(2014{\natexlab{b}})}]{spohn2014fluctuating}
\bibinfo{author}{\bibfnamefont{H.}~\bibnamefont{Spohn}},
  \bibinfo{journal}{arXiv preprint arXiv:1411.3907}
  (\bibinfo{year}{2014}{\natexlab{b}}).

\bibitem[{\citenamefont{Kulkarni et~al.}(2015)\citenamefont{Kulkarni, Huse, and
  Spohn}}]{Kulkarni2015}
\bibinfo{author}{\bibfnamefont{M.}~\bibnamefont{Kulkarni}},
  \bibinfo{author}{\bibfnamefont{D.~A.} \bibnamefont{Huse}}, \bibnamefont{and}
  \bibinfo{author}{\bibfnamefont{H.}~\bibnamefont{Spohn}},
  \bibinfo{journal}{Phys. Rev. A} \textbf{\bibinfo{volume}{92}},
  \bibinfo{pages}{043612} (\bibinfo{year}{2015}).

\bibitem[{\citenamefont{Xiong and Zhang}(2018)}]{Xiong2018}
\bibinfo{author}{\bibfnamefont{D.}~\bibnamefont{Xiong}} \bibnamefont{and}
  \bibinfo{author}{\bibfnamefont{Y.}~\bibnamefont{Zhang}},
  \bibinfo{journal}{Phys. Rev. E} \textbf{\bibinfo{volume}{98}},
  \bibinfo{pages}{012130} (\bibinfo{year}{2018}).

\bibitem[{\citenamefont{Di~Cintio et~al.}(2018)\citenamefont{Di~Cintio, Iubini,
  Lepri, and Livi}}]{DiCintio2018}
\bibinfo{author}{\bibfnamefont{P.}~\bibnamefont{Di~Cintio}},
  \bibinfo{author}{\bibfnamefont{S.}~\bibnamefont{Iubini}},
  \bibinfo{author}{\bibfnamefont{S.}~\bibnamefont{Lepri}}, \bibnamefont{and}
  \bibinfo{author}{\bibfnamefont{R.}~\bibnamefont{Livi}},
  \bibinfo{journal}{Chaos, Solitons \& Fractals}
  \textbf{\bibinfo{volume}{117}}, \bibinfo{pages}{249} (\bibinfo{year}{2018}).

\bibitem[{\citenamefont{Theodorakopoulos and Peyrard}(1999)}]{Theod1999}
\bibinfo{author}{\bibfnamefont{N.}~\bibnamefont{Theodorakopoulos}}
  \bibnamefont{and} \bibinfo{author}{\bibfnamefont{M.}~\bibnamefont{Peyrard}},
  \bibinfo{journal}{Phys. Rev. Lett.} \textbf{\bibinfo{volume}{83}},
  \bibinfo{pages}{2293} (\bibinfo{year}{1999}).

\bibitem[{\citenamefont{Spohn}(2018)}]{Spohn2018}
\bibinfo{author}{\bibfnamefont{H.}~\bibnamefont{Spohn}},
  \bibinfo{journal}{Journal of Mathematical Physics}
  \textbf{\bibinfo{volume}{59}}, \bibinfo{pages}{091402}
  (\bibinfo{year}{2018}), \eprint{https://doi.org/10.1063/1.5018624},
  \urlprefix\url{https://doi.org/10.1063/1.5018624}.

\bibitem[{\citenamefont{Kundu and Dhar}(2016)}]{Kundu2016}
\bibinfo{author}{\bibfnamefont{A.}~\bibnamefont{Kundu}} \bibnamefont{and}
  \bibinfo{author}{\bibfnamefont{A.}~\bibnamefont{Dhar}},
  \bibinfo{journal}{Phys. Rev. E} \textbf{\bibinfo{volume}{94}},
  \bibinfo{pages}{062130} (\bibinfo{year}{2016}),
  \urlprefix\url{https://link.aps.org/doi/10.1103/PhysRevE.94.062130}.

\bibitem[{\citenamefont{Dhar et~al.}(2019{\natexlab{b}})\citenamefont{Dhar,
  Kundu, Lebowitz, and Scaramazza}}]{Dhar2018}
\bibinfo{author}{\bibfnamefont{A.}~\bibnamefont{Dhar}},
  \bibinfo{author}{\bibfnamefont{A.}~\bibnamefont{Kundu}},
  \bibinfo{author}{\bibfnamefont{J.~L.} \bibnamefont{Lebowitz}},
  \bibnamefont{and} \bibinfo{author}{\bibfnamefont{J.~A.}
  \bibnamefont{Scaramazza}}, \bibinfo{journal}{Journal of Statistical Physics}
  \textbf{\bibinfo{volume}{175}}, \bibinfo{pages}{1298}
  (\bibinfo{year}{2019}{\natexlab{b}}).

\bibitem[{\citenamefont{Tamaki et~al.}(2017)\citenamefont{Tamaki, Sasada, and
  Saito}}]{Tamaki2017}
\bibinfo{author}{\bibfnamefont{S.}~\bibnamefont{Tamaki}},
  \bibinfo{author}{\bibfnamefont{M.}~\bibnamefont{Sasada}}, \bibnamefont{and}
  \bibinfo{author}{\bibfnamefont{K.}~\bibnamefont{Saito}},
  \bibinfo{journal}{Physical review letters} \textbf{\bibinfo{volume}{119}},
  \bibinfo{pages}{110602} (\bibinfo{year}{2017}).

\bibitem[{\citenamefont{Saito and Sasada}(2018)}]{Saito2018}
\bibinfo{author}{\bibfnamefont{K.}~\bibnamefont{Saito}} \bibnamefont{and}
  \bibinfo{author}{\bibfnamefont{M.}~\bibnamefont{Sasada}},
  \bibinfo{journal}{Communications in Mathematical Physics}
  \textbf{\bibinfo{volume}{361}}, \bibinfo{pages}{951} (\bibinfo{year}{2018}).

\bibitem[{\citenamefont{Bouchet et~al.}(2010)\citenamefont{Bouchet, Gupta, and
  Mukamel}}]{Bouchet2010}
\bibinfo{author}{\bibfnamefont{F.}~\bibnamefont{Bouchet}},
  \bibinfo{author}{\bibfnamefont{S.}~\bibnamefont{Gupta}}, \bibnamefont{and}
  \bibinfo{author}{\bibfnamefont{D.}~\bibnamefont{Mukamel}},
  \bibinfo{journal}{Physica A: Statistical Mechanics and its Applications}
  \textbf{\bibinfo{volume}{389}}, \bibinfo{pages}{4389} (\bibinfo{year}{2010}).

\bibitem[{\citenamefont{Campa et~al.}(2009)\citenamefont{Campa, Dauxois, and
  Ruffo}}]{RuffoRev}
\bibinfo{author}{\bibfnamefont{A.}~\bibnamefont{Campa}},
  \bibinfo{author}{\bibfnamefont{T.}~\bibnamefont{Dauxois}}, \bibnamefont{and}
  \bibinfo{author}{\bibfnamefont{S.}~\bibnamefont{Ruffo}},
  \bibinfo{journal}{Phys. Rep.} \textbf{\bibinfo{volume}{480}},
  \bibinfo{pages}{57} (\bibinfo{year}{2009}).

\bibitem[{\citenamefont{Gupta and Casetti}(2016)}]{gupta2016surprises}
\bibinfo{author}{\bibfnamefont{S.}~\bibnamefont{Gupta}} \bibnamefont{and}
  \bibinfo{author}{\bibfnamefont{L.}~\bibnamefont{Casetti}},
  \bibinfo{journal}{New Journal of Physics} \textbf{\bibinfo{volume}{18}},
  \bibinfo{pages}{103051} (\bibinfo{year}{2016}).

\bibitem[{\citenamefont{de~Buyl et~al.}(2013)\citenamefont{de~Buyl, De~Ninno,
  Fanelli, Nardini, Patelli, and Piazza}}]{deBuyl2013}
\bibinfo{author}{\bibfnamefont{P.}~\bibnamefont{de~Buyl}},
  \bibinfo{author}{\bibfnamefont{G.}~\bibnamefont{De~Ninno}},
  \bibinfo{author}{\bibfnamefont{D.}~\bibnamefont{Fanelli}},
  \bibinfo{author}{\bibfnamefont{C.}~\bibnamefont{Nardini}},
  \bibinfo{author}{\bibfnamefont{A.}~\bibnamefont{Patelli}}, \bibnamefont{and}
  \bibinfo{author}{\bibfnamefont{Y.~Y.} \bibnamefont{Piazza},
  \bibfnamefont{Francesco an d~Yamaguchi}}, \bibinfo{journal}{Phys. Rev. E}
  \textbf{\bibinfo{volume}{87}}, \bibinfo{pages}{042110}
  (\bibinfo{year}{2013}).

\bibitem[{\citenamefont{Torcini and Lepri}(1997)}]{Torcini1997}
\bibinfo{author}{\bibfnamefont{A.}~\bibnamefont{Torcini}} \bibnamefont{and}
  \bibinfo{author}{\bibfnamefont{S.}~\bibnamefont{Lepri}},
  \bibinfo{journal}{Phys. Rev. E} \textbf{\bibinfo{volume}{55}},
  \bibinfo{pages}{R3805} (\bibinfo{year}{1997}).

\bibitem[{\citenamefont{M{\'e}tivier et~al.}(2014)\citenamefont{M{\'e}tivier,
  Bachelard, and Kastner}}]{Metivier2014}
\bibinfo{author}{\bibfnamefont{D.}~\bibnamefont{M{\'e}tivier}},
  \bibinfo{author}{\bibfnamefont{R.}~\bibnamefont{Bachelard}},
  \bibnamefont{and} \bibinfo{author}{\bibfnamefont{M.}~\bibnamefont{Kastner}},
  \bibinfo{journal}{Phys. Rev. Lett.} \textbf{\bibinfo{volume}{112}},
  \bibinfo{pages}{210601} (\bibinfo{year}{2014}).

\bibitem[{\citenamefont{{\'A}vila et~al.}(2015)\citenamefont{{\'A}vila,
  Pereira, and Teixeira}}]{avila2015length}
\bibinfo{author}{\bibfnamefont{R.~R.} \bibnamefont{{\'A}vila}},
  \bibinfo{author}{\bibfnamefont{E.}~\bibnamefont{Pereira}}, \bibnamefont{and}
  \bibinfo{author}{\bibfnamefont{D.~L.} \bibnamefont{Teixeira}},
  \bibinfo{journal}{Physica A: Statistical Mechanics and its Applications}
  \textbf{\bibinfo{volume}{423}}, \bibinfo{pages}{51} (\bibinfo{year}{2015}).

\bibitem[{\citenamefont{Olivares and Anteneodo}(2016)}]{Olivares2016}
\bibinfo{author}{\bibfnamefont{C.}~\bibnamefont{Olivares}} \bibnamefont{and}
  \bibinfo{author}{\bibfnamefont{C.}~\bibnamefont{Anteneodo}},
  \bibinfo{journal}{Phys. Rev. E} \textbf{\bibinfo{volume}{94}},
  \bibinfo{pages}{042117} (\bibinfo{year}{2016}).

\bibitem[{\citenamefont{Bagchi}(2017)}]{Bagchi2017}
\bibinfo{author}{\bibfnamefont{D.}~\bibnamefont{Bagchi}},
  \bibinfo{journal}{Phys. Rev. E} \textbf{\bibinfo{volume}{95}},
  \bibinfo{pages}{032102} (\bibinfo{year}{2017}).

\bibitem[{\citenamefont{Tamaki and Saito}(2019)}]{tamaki2019energy}
\bibinfo{author}{\bibfnamefont{S.}~\bibnamefont{Tamaki}} \bibnamefont{and}
  \bibinfo{author}{\bibfnamefont{K.}~\bibnamefont{Saito}},
  \bibinfo{journal}{arXiv preprint arXiv:1906.08457}  (\bibinfo{year}{2019}).

\bibitem[{\citenamefont{Wang et~al.}(2019)\citenamefont{Wang, Dmitriev, and
  Xiong}}]{wang2019extremely}
\bibinfo{author}{\bibfnamefont{J.}~\bibnamefont{Wang}},
  \bibinfo{author}{\bibfnamefont{S.~V.} \bibnamefont{Dmitriev}},
  \bibnamefont{and} \bibinfo{author}{\bibfnamefont{D.}~\bibnamefont{Xiong}},
  \bibinfo{journal}{arXiv preprint arXiv:1906.11086}  (\bibinfo{year}{2019}).

\bibitem[{\citenamefont{Iubini et~al.}(2018)\citenamefont{Iubini, Di~Cintio,
  Lepri, Livi, and Casetti}}]{Iubini2018}
\bibinfo{author}{\bibfnamefont{S.}~\bibnamefont{Iubini}},
  \bibinfo{author}{\bibfnamefont{P.}~\bibnamefont{Di~Cintio}},
  \bibinfo{author}{\bibfnamefont{S.}~\bibnamefont{Lepri}},
  \bibinfo{author}{\bibfnamefont{R.}~\bibnamefont{Livi}}, \bibnamefont{and}
  \bibinfo{author}{\bibfnamefont{L.}~\bibnamefont{Casetti}},
  \bibinfo{journal}{Phys. Rev. E} \textbf{\bibinfo{volume}{97}},
  \bibinfo{pages}{032102} (\bibinfo{year}{2018}).

\bibitem[{\citenamefont{Di~Cintio et~al.}(2019)\citenamefont{Di~Cintio, Iubini,
  Lepri, and Livi}}]{DiCintio2019}
\bibinfo{author}{\bibfnamefont{P.}~\bibnamefont{Di~Cintio}},
  \bibinfo{author}{\bibfnamefont{S.}~\bibnamefont{Iubini}},
  \bibinfo{author}{\bibfnamefont{S.}~\bibnamefont{Lepri}}, \bibnamefont{and}
  \bibinfo{author}{\bibfnamefont{R.}~\bibnamefont{Livi}}, \bibinfo{journal}{J.
  Phys. A: Math. and Theoretical} \textbf{\bibinfo{volume}{52}},
  \bibinfo{pages}{274001} (\bibinfo{year}{2019}),
  \urlprefix\url{https://doi.org/10.1088%2F1751-8121%2Fab22f7}.

\bibitem[{\citenamefont{Malevanets and Kapral}(1999)}]{Kapral1999}
\bibinfo{author}{\bibfnamefont{A.}~\bibnamefont{Malevanets}} \bibnamefont{and}
  \bibinfo{author}{\bibfnamefont{R.}~\bibnamefont{Kapral}},
  \bibinfo{journal}{J. Chem. Phys.} \textbf{\bibinfo{volume}{110}},
  \bibinfo{pages}{8605} (\bibinfo{year}{1999}).

\bibitem[{\citenamefont{Ciraolo et~al.}(2018)\citenamefont{Ciraolo, Bufferand,
  Di~Cintio, Ghendrih, Lepri, Livi, Marandet, Serre, Tamain, and
  Valentinuzzi}}]{Ciraolo2018}
\bibinfo{author}{\bibfnamefont{G.}~\bibnamefont{Ciraolo}},
  \bibinfo{author}{\bibfnamefont{H.}~\bibnamefont{Bufferand}},
  \bibinfo{author}{\bibfnamefont{P.}~\bibnamefont{Di~Cintio}},
  \bibinfo{author}{\bibfnamefont{P.}~\bibnamefont{Ghendrih}},
  \bibinfo{author}{\bibfnamefont{S.}~\bibnamefont{Lepri}},
  \bibinfo{author}{\bibfnamefont{R.}~\bibnamefont{Livi}},
  \bibinfo{author}{\bibfnamefont{Y.}~\bibnamefont{Marandet}},
  \bibinfo{author}{\bibfnamefont{E.}~\bibnamefont{Serre}},
  \bibinfo{author}{\bibfnamefont{P.}~\bibnamefont{Tamain}}, \bibnamefont{and}
  \bibinfo{author}{\bibfnamefont{M.}~\bibnamefont{Valentinuzzi}},
  \bibinfo{journal}{Contributions to Plasma Physics}
  \textbf{\bibinfo{volume}{58}}, \bibinfo{pages}{457} (\bibinfo{year}{2018}).

\bibitem[{\citenamefont{Benenti et~al.}(2014)\citenamefont{Benenti, Casati, and
  Mej{\'\i}a-Monasterio}}]{Benenti2014}
\bibinfo{author}{\bibfnamefont{G.}~\bibnamefont{Benenti}},
  \bibinfo{author}{\bibfnamefont{G.}~\bibnamefont{Casati}}, \bibnamefont{and}
  \bibinfo{author}{\bibfnamefont{C.}~\bibnamefont{Mej{\'\i}a-Monasterio}},
  \bibinfo{journal}{New J. Phys.} \textbf{\bibinfo{volume}{16}},
  \bibinfo{pages}{015014} (\bibinfo{year}{2014}).

\bibitem[{\citenamefont{Di~Cintio et~al.}(2015)\citenamefont{Di~Cintio, Livi,
  Bufferand, Ciraolo, Lepri, and Straka}}]{DiCintio2015}
\bibinfo{author}{\bibfnamefont{P.}~\bibnamefont{Di~Cintio}},
  \bibinfo{author}{\bibfnamefont{R.}~\bibnamefont{Livi}},
  \bibinfo{author}{\bibfnamefont{H.}~\bibnamefont{Bufferand}},
  \bibinfo{author}{\bibfnamefont{G.}~\bibnamefont{Ciraolo}},
  \bibinfo{author}{\bibfnamefont{S.}~\bibnamefont{Lepri}}, \bibnamefont{and}
  \bibinfo{author}{\bibfnamefont{M.~J.} \bibnamefont{Straka}},
  \bibinfo{journal}{Phys. Rev. E} \textbf{\bibinfo{volume}{92}},
  \bibinfo{pages}{062108} (\bibinfo{year}{2015}).

\bibitem[{\citenamefont{Upadhyaya and Aksamija}(2016)}]{Upadhyaya2016}
\bibinfo{author}{\bibfnamefont{M.}~\bibnamefont{Upadhyaya}} \bibnamefont{and}
  \bibinfo{author}{\bibfnamefont{Z.}~\bibnamefont{Aksamija}},
  \bibinfo{journal}{Physical Review B} \textbf{\bibinfo{volume}{94}},
  \bibinfo{pages}{174303} (\bibinfo{year}{2016}).

\bibitem[{\citenamefont{Crnjar et~al.}(2018)\citenamefont{Crnjar, Melis, and
  Colombo}}]{Colombo2018}
\bibinfo{author}{\bibfnamefont{A.}~\bibnamefont{Crnjar}},
  \bibinfo{author}{\bibfnamefont{C.}~\bibnamefont{Melis}}, \bibnamefont{and}
  \bibinfo{author}{\bibfnamefont{L.}~\bibnamefont{Colombo}},
  \bibinfo{journal}{Phys. Rev. Materials} \textbf{\bibinfo{volume}{2}},
  \bibinfo{pages}{015603} (\bibinfo{year}{2018}).

\bibitem[{\citenamefont{Sun et~al.}(2015)\citenamefont{Sun, Zhang, Li, Xia,
  Zhang, Du, Isikgor, and Ouyang}}]{Sun2015}
\bibinfo{author}{\bibfnamefont{K.}~\bibnamefont{Sun}},
  \bibinfo{author}{\bibfnamefont{S.}~\bibnamefont{Zhang}},
  \bibinfo{author}{\bibfnamefont{P.}~\bibnamefont{Li}},
  \bibinfo{author}{\bibfnamefont{Y.}~\bibnamefont{Xia}},
  \bibinfo{author}{\bibfnamefont{X.}~\bibnamefont{Zhang}},
  \bibinfo{author}{\bibfnamefont{D.}~\bibnamefont{Du}},
  \bibinfo{author}{\bibfnamefont{F.~H.} \bibnamefont{Isikgor}},
  \bibnamefont{and} \bibinfo{author}{\bibfnamefont{J.}~\bibnamefont{Ouyang}},
  \bibinfo{journal}{Journal of Materials Science: Materials in Electronics}
  \textbf{\bibinfo{volume}{26}}, \bibinfo{pages}{4438} (\bibinfo{year}{2015}),
  ISSN \bibinfo{issn}{1573-482X},
  \urlprefix\url{https://doi.org/10.1007/s10854-015-2895-5}.

\bibitem[{\citenamefont{Barreto et~al.}(2019)\citenamefont{Barreto, Carusela,
  and Monastra}}]{Barreto2019}
\bibinfo{author}{\bibfnamefont{R.}~\bibnamefont{Barreto}},
  \bibinfo{author}{\bibfnamefont{M.}~\bibnamefont{Carusela}}, \bibnamefont{and}
  \bibinfo{author}{\bibfnamefont{A.}~\bibnamefont{Monastra}},
  \bibinfo{journal}{Phys. Rev. E} \textbf{\bibinfo{volume}{100}},
  \bibinfo{pages}{022118} (\bibinfo{year}{2019}).

\bibitem[{\citenamefont{Lee et~al.}(2017)\citenamefont{Lee, Wu, Lou, Lee, and
  Chang}}]{Lee2017}
\bibinfo{author}{\bibfnamefont{V.}~\bibnamefont{Lee}},
  \bibinfo{author}{\bibfnamefont{C.-H.} \bibnamefont{Wu}},
  \bibinfo{author}{\bibfnamefont{Z.-X.} \bibnamefont{Lou}},
  \bibinfo{author}{\bibfnamefont{W.-L.} \bibnamefont{Lee}}, \bibnamefont{and}
  \bibinfo{author}{\bibfnamefont{C.-W.} \bibnamefont{Chang}},
  \bibinfo{journal}{Physical review letters} \textbf{\bibinfo{volume}{118}},
  \bibinfo{pages}{135901} (\bibinfo{year}{2017}).

\bibitem[{\citenamefont{Li et~al.}(2017)\citenamefont{Li, Takahashi, and
  Zhang}}]{Li2017comment}
\bibinfo{author}{\bibfnamefont{Q.-Y.} \bibnamefont{Li}},
  \bibinfo{author}{\bibfnamefont{K.}~\bibnamefont{Takahashi}},
  \bibnamefont{and} \bibinfo{author}{\bibfnamefont{X.}~\bibnamefont{Zhang}},
  \bibinfo{journal}{Physical review letters} \textbf{\bibinfo{volume}{119}},
  \bibinfo{pages}{179601} (\bibinfo{year}{2017}).

\bibitem[{\citenamefont{Meier et~al.}(2014)\citenamefont{Meier, Menges,
  Nirmalraj, H{\"o}lscher, Riel, and Gotsmann}}]{Meier2014}
\bibinfo{author}{\bibfnamefont{T.}~\bibnamefont{Meier}},
  \bibinfo{author}{\bibfnamefont{F.}~\bibnamefont{Menges}},
  \bibinfo{author}{\bibfnamefont{P.}~\bibnamefont{Nirmalraj}},
  \bibinfo{author}{\bibfnamefont{H.}~\bibnamefont{H{\"o}lscher}},
  \bibinfo{author}{\bibfnamefont{H.}~\bibnamefont{Riel}}, \bibnamefont{and}
  \bibinfo{author}{\bibfnamefont{B.}~\bibnamefont{Gotsmann}},
  \bibinfo{journal}{Phys. Rev. Lett.} \textbf{\bibinfo{volume}{113}},
  \bibinfo{pages}{060801} (\bibinfo{year}{2014}).

\bibitem[{\citenamefont{Benenti et~al.}(2016)\citenamefont{Benenti, Casati,
  Mej{\'\i}a-Monasterio, and M.}}]{Benenti2016}
\bibinfo{author}{\bibfnamefont{G.}~\bibnamefont{Benenti}},
  \bibinfo{author}{\bibfnamefont{G.}~\bibnamefont{Casati}},
  \bibinfo{author}{\bibfnamefont{C.}~\bibnamefont{Mej{\'\i}a-Monasterio}},
  \bibnamefont{and} \bibinfo{author}{\bibfnamefont{P.}~\bibnamefont{M.}},
  \emph{\bibinfo{title}{From thermal rectifiers to thermoelectric devices,}}
  (\bibinfo{publisher}{Springer}, \bibinfo{year}{2016}),
  chap.~\bibinfo{chapter}{10}.

\bibitem[{\citenamefont{Andresen}(2011)}]{Andresen2011}
\bibinfo{author}{\bibfnamefont{B.}~\bibnamefont{Andresen}},
  \bibinfo{journal}{Angew. Chem. Int. Ed.}  (\bibinfo{year}{2011}).

\bibitem[{\citenamefont{Shiraishi et~al.}(2016)\citenamefont{Shiraishi, Saito,
  and Tasaki}}]{Saito2016}
\bibinfo{author}{\bibfnamefont{N.}~\bibnamefont{Shiraishi}},
  \bibinfo{author}{\bibfnamefont{K.}~\bibnamefont{Saito}}, \bibnamefont{and}
  \bibinfo{author}{\bibfnamefont{H.}~\bibnamefont{Tasaki}},
  \bibinfo{journal}{Phys. Rev. Lett.} \textbf{\bibinfo{volume}{117}},
  \bibinfo{pages}{190601} (\bibinfo{year}{2016}).

\bibitem[{\citenamefont{Brandner et~al.}(2015)\citenamefont{Brandner, Saito,
  and Seifert}}]{Seifert2015}
\bibinfo{author}{\bibfnamefont{K.}~\bibnamefont{Brandner}},
  \bibinfo{author}{\bibfnamefont{K.}~\bibnamefont{Saito}}, \bibnamefont{and}
  \bibinfo{author}{\bibfnamefont{U.}~\bibnamefont{Seifert}},
  \bibinfo{journal}{Phys. Rev. X} \textbf{\bibinfo{volume}{5}},
  \bibinfo{pages}{031019} (\bibinfo{year}{2015}).

\bibitem[{\citenamefont{Seifert}(2012)}]{Seifert2012}
\bibinfo{author}{\bibfnamefont{U.}~\bibnamefont{Seifert}},
  \bibinfo{journal}{Rep. Prog. Phys.} \textbf{\bibinfo{volume}{75}},
  \bibinfo{pages}{126001} (\bibinfo{year}{2012}).

\bibitem[{\citenamefont{Campisi and Fazio}(2016)}]{Campisi2016}
\bibinfo{author}{\bibfnamefont{M.}~\bibnamefont{Campisi}} \bibnamefont{and}
  \bibinfo{author}{\bibfnamefont{R.}~\bibnamefont{Fazio}},
  \bibinfo{journal}{Nat. Commun.} \textbf{\bibinfo{volume}{7}},
  \bibinfo{pages}{11895} (\bibinfo{year}{2016}).

\bibitem[{\citenamefont{Pietzonka and Seifert}(2018)}]{Seifert2018}
\bibinfo{author}{\bibfnamefont{P.}~\bibnamefont{Pietzonka}} \bibnamefont{and}
  \bibinfo{author}{\bibfnamefont{U.}~\bibnamefont{Seifert}},
  \bibinfo{journal}{Phys. Rev. Lett.} \textbf{\bibinfo{volume}{120}},
  \bibinfo{pages}{190602} (\bibinfo{year}{2018}).

\bibitem[{\citenamefont{Koyuk and Seifert}(2019)}]{Seifert2019}
\bibinfo{author}{\bibfnamefont{T.}~\bibnamefont{Koyuk}} \bibnamefont{and}
  \bibinfo{author}{\bibfnamefont{U.}~\bibnamefont{Seifert}},
  \bibinfo{journal}{Phys. Rev. Lett.} \textbf{\bibinfo{volume}{122}},
  \bibinfo{pages}{230601} (\bibinfo{year}{2019}).

\bibitem[{\citenamefont{Callen}(1985)}]{callen}
\bibinfo{author}{\bibfnamefont{H.~B.} \bibnamefont{Callen}},
  \emph{\bibinfo{title}{Thermodynamics and an introduction to thermostatics}}
  (\bibinfo{publisher}{John Wiley \& Sons, New York}, \bibinfo{year}{1985}).

\bibitem[{\citenamefont{de~Groot and Mazur}(1984)}]{mazur}
\bibinfo{author}{\bibfnamefont{S.~R.} \bibnamefont{de~Groot}} \bibnamefont{and}
  \bibinfo{author}{\bibfnamefont{P.}~\bibnamefont{Mazur}},
  \emph{\bibinfo{title}{Non-equilibrium thermodynamics}}
  (\bibinfo{publisher}{Dover, New York}, \bibinfo{year}{1984}).

\bibitem[{\citenamefont{Zotos et~al.}(1997)\citenamefont{Zotos, Naef, and
  Prelov\v{s}ek}}]{Zotos1997}
\bibinfo{author}{\bibfnamefont{X.}~\bibnamefont{Zotos}},
  \bibinfo{author}{\bibfnamefont{F.~P.} \bibnamefont{Naef}}, \bibnamefont{and}
  \bibinfo{author}{\bibfnamefont{P.}~\bibnamefont{Prelov\v{s}ek}},
  \bibinfo{journal}{Phys. Rev. E} \textbf{\bibinfo{volume}{55}},
  \bibinfo{pages}{11029} (\bibinfo{year}{1997}).

\bibitem[{\citenamefont{Zotos and Prelov\v{s}ek}(2004)}]{Zotos2004}
\bibinfo{author}{\bibfnamefont{X.}~\bibnamefont{Zotos}} \bibnamefont{and}
  \bibinfo{author}{\bibfnamefont{P.}~\bibnamefont{Prelov\v{s}ek}},
  \emph{\bibinfo{title}{Transport in one dimensional quantum systems}}
  (\bibinfo{publisher}{Kluwer Academic Piublishers}, \bibinfo{year}{2004}),
  chap.~\bibinfo{chapter}{11}.

\bibitem[{\citenamefont{Garst and Rosch}(2001)}]{Garst2001}
\bibinfo{author}{\bibfnamefont{M.}~\bibnamefont{Garst}} \bibnamefont{and}
  \bibinfo{author}{\bibfnamefont{A.}~\bibnamefont{Rosch}},
  \bibinfo{journal}{EPL (Europhysics Letters)} \textbf{\bibinfo{volume}{55}},
  \bibinfo{pages}{66} (\bibinfo{year}{2001}).

\bibitem[{\citenamefont{Heidrich-Meisner
  et~al.}(2005)\citenamefont{Heidrich-Meisner, Hoenecker, and
  Brenig}}]{Heidrichmeisner2005}
\bibinfo{author}{\bibfnamefont{F.}~\bibnamefont{Heidrich-Meisner}},
  \bibinfo{author}{\bibfnamefont{A.}~\bibnamefont{Hoenecker}},
  \bibnamefont{and} \bibinfo{author}{\bibfnamefont{W.}~\bibnamefont{Brenig}},
  \bibinfo{journal}{Phys. Rev. B} \textbf{\bibinfo{volume}{71}},
  \bibinfo{pages}{184415} (\bibinfo{year}{2005}).

\bibitem[{\citenamefont{Mazur}(1969)}]{Mazur1969}
\bibinfo{author}{\bibfnamefont{P.}~\bibnamefont{Mazur}},
  \bibinfo{journal}{Physica} \textbf{\bibinfo{volume}{43}},
  \bibinfo{pages}{533} (\bibinfo{year}{1969}).

\bibitem[{\citenamefont{Suzuki}(1971)}]{Suzuki1971}
\bibinfo{author}{\bibfnamefont{M.}~\bibnamefont{Suzuki}},
  \bibinfo{journal}{Physica} \textbf{\bibinfo{volume}{51}},
  \bibinfo{pages}{277} (\bibinfo{year}{1971}).

\bibitem[{\citenamefont{Benenti et~al.}(2013)\citenamefont{Benenti, Casati, and
  Wang}}]{Benenti2013}
\bibinfo{author}{\bibfnamefont{G.}~\bibnamefont{Benenti}},
  \bibinfo{author}{\bibfnamefont{G.}~\bibnamefont{Casati}}, \bibnamefont{and}
  \bibinfo{author}{\bibfnamefont{J.}~\bibnamefont{Wang}},
  \bibinfo{journal}{Phys. Rev. Lett.} \textbf{\bibinfo{volume}{110}},
  \bibinfo{pages}{70604} (\bibinfo{year}{2013}).

\bibitem[{\citenamefont{Luo et~al.}(2018)\citenamefont{Luo, Benenti, Casati,
  and Wang}}]{Luo2018}
\bibinfo{author}{\bibfnamefont{R.}~\bibnamefont{Luo}},
  \bibinfo{author}{\bibfnamefont{G.}~\bibnamefont{Benenti}},
  \bibinfo{author}{\bibfnamefont{G.}~\bibnamefont{Casati}}, \bibnamefont{and}
  \bibinfo{author}{\bibfnamefont{J.}~\bibnamefont{Wang}},
  \bibinfo{journal}{Phys. Rev. Lett.} \textbf{\bibinfo{volume}{121}},
  \bibinfo{pages}{080602} (\bibinfo{year}{2018}).

\bibitem[{\citenamefont{Chen et~al.}(2015)\citenamefont{Chen, Wang, Casati, and
  Benenti}}]{Chen2015}
\bibinfo{author}{\bibfnamefont{S.}~\bibnamefont{Chen}},
  \bibinfo{author}{\bibfnamefont{J.}~\bibnamefont{Wang}},
  \bibinfo{author}{\bibfnamefont{G.}~\bibnamefont{Casati}}, \bibnamefont{and}
  \bibinfo{author}{\bibfnamefont{G.}~\bibnamefont{Benenti}},
  \bibinfo{journal}{Phys. Rev. E} \textbf{\bibinfo{volume}{92}},
  \bibinfo{pages}{032139} (\bibinfo{year}{2015}).

\bibitem[{\citenamefont{Mej\'ia-Monasterio
  et~al.}(2001)\citenamefont{Mej\'ia-Monasterio, Larralde, and
  Leyvraz}}]{Mejia2001}
\bibinfo{author}{\bibfnamefont{C.}~\bibnamefont{Mej\'ia-Monasterio}},
  \bibinfo{author}{\bibfnamefont{H.}~\bibnamefont{Larralde}}, \bibnamefont{and}
  \bibinfo{author}{\bibfnamefont{F.}~\bibnamefont{Leyvraz}},
  \bibinfo{journal}{Phys. Rev. Lett.} \textbf{\bibinfo{volume}{86}},
  \bibinfo{pages}{5417} (\bibinfo{year}{2001}).

\bibitem[{\citenamefont{Larralde et~al.}(2003)\citenamefont{Larralde, Leyvraz,
  and Mej\'ia-Monasterio}}]{Larralde03}
\bibinfo{author}{\bibfnamefont{H.}~\bibnamefont{Larralde}},
  \bibinfo{author}{\bibfnamefont{F.}~\bibnamefont{Leyvraz}}, \bibnamefont{and}
  \bibinfo{author}{\bibfnamefont{C.}~\bibnamefont{Mej\'ia-Monasterio}},
  \bibinfo{journal}{J. Stat. Phys.} \textbf{\bibinfo{volume}{113}},
  \bibinfo{pages}{197} (\bibinfo{year}{2003}).

\bibitem[{\citenamefont{Datta}(1995)}]{datta}
\bibinfo{author}{\bibfnamefont{S.}~\bibnamefont{Datta}},
  \emph{\bibinfo{title}{Electronic transport in mesoscopic systems}}
  (\bibinfo{publisher}{Cambridge University Press}, \bibinfo{year}{1995}).

\bibitem[{\citenamefont{Whitney}(2014)}]{Whitney2014}
\bibinfo{author}{\bibfnamefont{R.~S.} \bibnamefont{Whitney}},
  \bibinfo{journal}{Phys. Rev. Lett.} \textbf{\bibinfo{volume}{112}},
  \bibinfo{pages}{130601} (\bibinfo{year}{2014}).

\bibitem[{\citenamefont{Whitney}(2015)}]{Whitney2015}
\bibinfo{author}{\bibfnamefont{R.~S.} \bibnamefont{Whitney}},
  \bibinfo{journal}{Phys. Rev. B} \textbf{\bibinfo{volume}{91}},
  \bibinfo{pages}{115425} (\bibinfo{year}{2015}).

\bibitem[{\citenamefont{Mahan and Sofo}(1996)}]{Mahan1996}
\bibinfo{author}{\bibfnamefont{G.~D.} \bibnamefont{Mahan}} \bibnamefont{and}
  \bibinfo{author}{\bibfnamefont{J.~O.} \bibnamefont{Sofo}},
  \bibinfo{journal}{PNAS} \textbf{\bibinfo{volume}{93}}, \bibinfo{pages}{7436}
  (\bibinfo{year}{1996}).

\bibitem[{\citenamefont{Humphrey et~al.}(2002)\citenamefont{Humphrey, Newbury,
  Taylor, and Linke}}]{Humphrey2002}
\bibinfo{author}{\bibfnamefont{T.~E.} \bibnamefont{Humphrey}},
  \bibinfo{author}{\bibfnamefont{R.}~\bibnamefont{Newbury}},
  \bibinfo{author}{\bibfnamefont{R.~P.} \bibnamefont{Taylor}},
  \bibnamefont{and} \bibinfo{author}{\bibfnamefont{H.}~\bibnamefont{Linke}},
  \bibinfo{journal}{Phys. Rev. Lett.} \textbf{\bibinfo{volume}{89}},
  \bibinfo{pages}{116801} (\bibinfo{year}{2002}).

\bibitem[{\citenamefont{Humphrey and Linke}(2005)}]{Humphrey2005}
\bibinfo{author}{\bibfnamefont{T.~E.} \bibnamefont{Humphrey}} \bibnamefont{and}
  \bibinfo{author}{\bibfnamefont{H.}~\bibnamefont{Linke}},
  \bibinfo{journal}{Phys. Rev. Lett.} \textbf{\bibinfo{volume}{94}},
  \bibinfo{pages}{096601} (\bibinfo{year}{2005}).

\bibitem[{\citenamefont{Saito et~al.}(2010)\citenamefont{Saito, Benenti, and
  Casati}}]{Saito2010}
\bibinfo{author}{\bibfnamefont{K.}~\bibnamefont{Saito}},
  \bibinfo{author}{\bibfnamefont{G.}~\bibnamefont{Benenti}}, \bibnamefont{and}
  \bibinfo{author}{\bibfnamefont{G.}~\bibnamefont{Casati}},
  \bibinfo{journal}{Chem. Phys.} \textbf{\bibinfo{volume}{375}},
  \bibinfo{pages}{508} (\bibinfo{year}{2010}).

\bibitem[{\citenamefont{Brantut et~al.}(2013)\citenamefont{Brantut, Grenier,
  Meineke, Stadler, Krinner, Kollath, Esslinger, and Georges}}]{Brantut2013}
\bibinfo{author}{\bibfnamefont{J.-P.} \bibnamefont{Brantut}},
  \bibinfo{author}{\bibfnamefont{C.}~\bibnamefont{Grenier}},
  \bibinfo{author}{\bibfnamefont{J.}~\bibnamefont{Meineke}},
  \bibinfo{author}{\bibfnamefont{D.}~\bibnamefont{Stadler}},
  \bibinfo{author}{\bibfnamefont{S.}~\bibnamefont{Krinner}},
  \bibinfo{author}{\bibfnamefont{C.}~\bibnamefont{Kollath}},
  \bibinfo{author}{\bibfnamefont{T.}~\bibnamefont{Esslinger}},
  \bibnamefont{and} \bibinfo{author}{\bibfnamefont{A.}~\bibnamefont{Georges}},
  \bibinfo{journal}{Science} \textbf{\bibinfo{volume}{342}},
  \bibinfo{pages}{713} (\bibinfo{year}{2013}).

\bibitem[{\citenamefont{Husmann et~al.}(2018)\citenamefont{Husmann, Lebrat,
  H\"{a}usler, Brantut, Corman, and Esslinger}}]{Husmann2018}
\bibinfo{author}{\bibfnamefont{D.}~\bibnamefont{Husmann}},
  \bibinfo{author}{\bibfnamefont{M.}~\bibnamefont{Lebrat}},
  \bibinfo{author}{\bibfnamefont{S.}~\bibnamefont{H\"{a}usler}},
  \bibinfo{author}{\bibfnamefont{J.-P.} \bibnamefont{Brantut}},
  \bibinfo{author}{\bibfnamefont{L.}~\bibnamefont{Corman}}, \bibnamefont{and}
  \bibinfo{author}{\bibfnamefont{T.}~\bibnamefont{Esslinger}},
  \bibinfo{journal}{PNAS} \textbf{\bibinfo{volume}{115}}, \bibinfo{pages}{8563}
  (\bibinfo{year}{2018}).

\bibitem[{\citenamefont{Vasseur and Moore}(2016)}]{Moore2016}
\bibinfo{author}{\bibfnamefont{R.}~\bibnamefont{Vasseur}} \bibnamefont{and}
  \bibinfo{author}{\bibfnamefont{J.~E.} \bibnamefont{Moore}},
  \bibinfo{journal}{J. Stat. Mech.} p. \bibinfo{pages}{064010}
  (\bibinfo{year}{2016}).

\bibitem[{\citenamefont{Abanin et~al.}(2019)\citenamefont{Abanin, Altman,
  Bloch, and Serbyn}}]{Abanin2019}
\bibinfo{author}{\bibfnamefont{D.~A.} \bibnamefont{Abanin}},
  \bibinfo{author}{\bibfnamefont{E.}~\bibnamefont{Altman}},
  \bibinfo{author}{\bibfnamefont{I.}~\bibnamefont{Bloch}}, \bibnamefont{and}
  \bibinfo{author}{\bibfnamefont{M.}~\bibnamefont{Serbyn}},
  \bibinfo{journal}{Rev. Mod. Phys.} \textbf{\bibinfo{volume}{91}},
  \bibinfo{pages}{021001} (\bibinfo{year}{2019}).

\bibitem[{\citenamefont{Prosen and Znidari{\v c}}(2009)}]{Prosen2009}
\bibinfo{author}{\bibfnamefont{T.}~\bibnamefont{Prosen}} \bibnamefont{and}
  \bibinfo{author}{\bibfnamefont{M.}~\bibnamefont{Znidari{\v c}}},
  \bibinfo{journal}{J. Stat. Mech.} p. \bibinfo{pages}{P02035}
  (\bibinfo{year}{2009}).

\bibitem[{\citenamefont{Mejia-Monasterio and
  Wichterich}(2007)}]{Mejia-Monasterio2007}
\bibinfo{author}{\bibfnamefont{C.}~\bibnamefont{Mejia-Monasterio}}
  \bibnamefont{and}
  \bibinfo{author}{\bibfnamefont{H.}~\bibnamefont{Wichterich}},
  \bibinfo{journal}{The European Physical Journal Special Topics}
  \textbf{\bibinfo{volume}{151}}, \bibinfo{pages}{113} (\bibinfo{year}{2007}).

\bibitem[{\citenamefont{Saito}(2003)}]{Saito2003}
\bibinfo{author}{\bibfnamefont{K.}~\bibnamefont{Saito}},
  \bibinfo{journal}{Europhys. Lett.} \textbf{\bibinfo{volume}{61}},
  \bibinfo{pages}{34} (\bibinfo{year}{2003}).

\bibitem[{\citenamefont{Mej\'{\i}a-Monasterio
  et~al.}(2005)\citenamefont{Mej\'{\i}a-Monasterio, Casati, and
  Prosen}}]{Monasterio2005}
\bibinfo{author}{\bibfnamefont{C.}~\bibnamefont{Mej\'{\i}a-Monasterio}},
  \bibinfo{author}{\bibfnamefont{G.}~\bibnamefont{Casati}}, \bibnamefont{and}
  \bibinfo{author}{\bibfnamefont{T.}~\bibnamefont{Prosen}},
  \bibinfo{journal}{Europhys. Lett.} \textbf{\bibinfo{volume}{72}},
  \bibinfo{pages}{520} (\bibinfo{year}{2005}).

\bibitem[{\citenamefont{Znidari{\v c} et~al.}(2010)\citenamefont{Znidari{\v c},
  Prosen, Benenti, Casati, and Rossini}}]{Znidaric2010}
\bibinfo{author}{\bibfnamefont{M.}~\bibnamefont{Znidari{\v c}}},
  \bibinfo{author}{\bibfnamefont{T.}~\bibnamefont{Prosen}},
  \bibinfo{author}{\bibfnamefont{G.}~\bibnamefont{Benenti}},
  \bibinfo{author}{\bibfnamefont{G.}~\bibnamefont{Casati}}, \bibnamefont{and}
  \bibinfo{author}{\bibfnamefont{D.}~\bibnamefont{Rossini}},
  \bibinfo{journal}{Phys. Rev. E} \textbf{\bibinfo{volume}{81}},
  \bibinfo{pages}{051135} (\bibinfo{year}{2010}).

\bibitem[{\citenamefont{Levy and Kosloff}(2014)}]{Levy2014}
\bibinfo{author}{\bibfnamefont{A.}~\bibnamefont{Levy}} \bibnamefont{and}
  \bibinfo{author}{\bibfnamefont{R.}~\bibnamefont{Kosloff}},
  \bibinfo{journal}{EPL} \textbf{\bibinfo{volume}{107}}, \bibinfo{pages}{2004}
  (\bibinfo{year}{2014}).

\bibitem[{\citenamefont{Balachandran et~al.}(2019)\citenamefont{Balachandran,
  Benenti, Pereira, Casati, and Poletti}}]{Balachandran2019}
\bibinfo{author}{\bibfnamefont{V.}~\bibnamefont{Balachandran}},
  \bibinfo{author}{\bibfnamefont{G.}~\bibnamefont{Benenti}},
  \bibinfo{author}{\bibfnamefont{E.}~\bibnamefont{Pereira}},
  \bibinfo{author}{\bibfnamefont{G.}~\bibnamefont{Casati}}, \bibnamefont{and}
  \bibinfo{author}{\bibfnamefont{D.}~\bibnamefont{Poletti}},
  \bibinfo{journal}{Phys. Rev. E} \textbf{\bibinfo{volume}{99}},
  \bibinfo{pages}{032136} (\bibinfo{year}{2019}).

\bibitem[{\citenamefont{Carrega et~al.}(2015)\citenamefont{Carrega, Solinas,
  Braggio, Sassetti, and Weiss}}]{Carrega2015}
\bibinfo{author}{\bibfnamefont{M.}~\bibnamefont{Carrega}},
  \bibinfo{author}{\bibfnamefont{P.}~\bibnamefont{Solinas}},
  \bibinfo{author}{\bibfnamefont{A.}~\bibnamefont{Braggio}},
  \bibinfo{author}{\bibfnamefont{M.}~\bibnamefont{Sassetti}}, \bibnamefont{and}
  \bibinfo{author}{\bibfnamefont{U.}~\bibnamefont{Weiss}},
  \bibinfo{journal}{New J. Phys.}  (\bibinfo{year}{2015}).

\bibitem[{\citenamefont{Tamascelli et~al.}(2019)\citenamefont{Tamascelli,
  Smirne, Lim, Huelga, and Plenio}}]{Plenio2019}
\bibinfo{author}{\bibfnamefont{D.}~\bibnamefont{Tamascelli}},
  \bibinfo{author}{\bibfnamefont{A.}~\bibnamefont{Smirne}},
  \bibinfo{author}{\bibfnamefont{J.}~\bibnamefont{Lim}},
  \bibinfo{author}{\bibfnamefont{S.~F.} \bibnamefont{Huelga}},
  \bibnamefont{and} \bibinfo{author}{\bibfnamefont{M.~B.}
  \bibnamefont{Plenio}}, \bibinfo{journal}{Phys. Rev. Lett.}
  \textbf{\bibinfo{volume}{123}}, \bibinfo{pages}{090402}
  (\bibinfo{year}{2019}).

\bibitem[{\citenamefont{Wiedmann et~al.}(2019)\citenamefont{Wiedmann,
  Stockburger, and Ankerjold}}]{Ankerhold2019}
\bibinfo{author}{\bibfnamefont{M.}~\bibnamefont{Wiedmann}},
  \bibinfo{author}{\bibfnamefont{T.}~\bibnamefont{Stockburger},
  \bibfnamefont{J}}, \bibnamefont{and}
  \bibinfo{author}{\bibfnamefont{J.}~\bibnamefont{Ankerjold}},
  \bibinfo{journal}{arXiv:1903.11368}  (\bibinfo{year}{2019}).

\end{thebibliography}
\bibliographystyle{apsrev}

\end{document}